\documentclass{revtex4}

\usepackage{amssymb}

\usepackage{color}
\definecolor{red}{rgb}{1,0,0}

\definecolor{green}{rgb}{0,1,1}

\def\giorno{8/09/2013}

\def\a{\alpha}
\def\b{\beta}

\def\ga{\gamma}
\def\de{\delta}   
\def\eps{\varepsilon}
\def\phi{\varphi}
\def\la{\lambda}

\def\s{\sigma}

\def\om{\omega}
\def\th{\theta}

\def\H{{\bf H}}

\def\h{{\cal H}}

\def\Jb{{\bf J}}

\def\L{{\cal L}}

\def\P{{\cal P}}
\def\Q{{\cal Q}}
\def\R{{\bf R}}
\def\S{{\cal S}}
\def\T{{\rm T}}

\def\Ga{\Gamma}
\def\De{\Delta}
\def\La{\Lambda}
\def\Om{\Omega}

\def\pa{\partial}

\def\d{{\rm d}}       
\def\w{\wedge}

\def\o+{\oplus}

\def\grad{\nabla}     
\def\ss{\subset}
\def\sse{\subseteq}

\def\<{\langle}
\def\>{\rangle}

\def\EOR{\hfill {$\odot$}}
\def\EOP{\hfill {$\triangle$}}

\def\interno{\hskip 2pt \vbox{\hbox{\vbox to .18
truecm{\vfill\hbox to .25 truecm
{\hfill\hfill}\vfill}\vrule}\hrule}\hskip 2 pt}

\def\({\left(}
\def\){\right)}
\def\[{\left[}
\def\]{\right]}
\def\=#1{\overline #1}
\def\ol#1{\overline #1}
\def\~#1{\widetilde #1}
\def\wt#1{\widetilde #1}
\def\.#1{\dot #1}
\def\^#1{\widehat #1}

\def\beq{\begin{equation}}
\def\eeq{\end{equation}}
\def\eqref#1{(\ref{#1})}

\def\salta#1{{}}

\def\hK{hyperkahler }
\def\hS{hypersymplectic }

\def\hSp{\mathtt{hSp}}

\def\Det{\mathtt{Det}}

\begin{document}

\title[Canonical transformations for HK structures]{Canonical transformations for hyperkahler structures and hyperhamiltonian dynamics}

\author{G. Gaeta$^1$\footnote{Electronic mail: giuseppe.gaeta@unimi.it} and M. A. Rodr\'{\i}guez$^2$\footnote{Electronic mail: rodrigue@fis.ucm.es}}

\affiliation{$^1$Dipartimento di Matematica, Universit\`a degli Studi di Milano, via Saldini 50, 20133 Milano (Italy)\\ $^2$Departamento de F\'{\i}sica Te\'orica II, Universidad Complutense, 28040 Madrid (Spain)}

\hfill {\tt Version \ of \ \giorno}

\begin{abstract}
We discuss generalizations of the well known concept of canonical
transformations for symplectic structures to the case of
hyperkahler structures. Different characterizations, which are
equivalent in the symplectic case, give raise to non-equivalent
notions in the hyperkahler framework; we will thus distinguish
between hyperkahler and canonical transformations. We also discuss
the properties of hyperhamiltonian dynamics in this respect.

\end{abstract}

\maketitle


\section{Introduction}

Following the pioneering work by Atiyah, Hitchin and their
coworkers \cite{At1,At2,AtH,Hit,HKLR}, hyperkahler manifolds and
structures were recognized to be relevant not only to Geometry
\cite{Cal,Cal2} but also to Physics, in particular in the context
of Field Theory and in connection to instantons and their moduli
spaces \cite{AtH,AtMS,Che,ChH,ChK,Dan,Dun,DuM,Hit,Man,PeP}.

More recently, an extension of standard Hamilton dynamics based on
hyperkahler structures (and defined on hyperkahler manifolds) has
been formulated \cite{GM}; the physical motivation behind this was
an attempt to have a classical framework for spin dynamics. It was
recently shown, indeed, that several of the fundamental equations
for the dynamics of spin (Pauli equation and the Dirac equation,
the latter in both the Foldy-Wouthuysen and the Cini-Touschek
frameworks) can be cast in the framework of hyperhamiltonian
dynamics \cite{GR}. This parallel with Hamiltonian dynamics calls
naturally for an extension (if possible) of the concepts and
constructions which are at the roots of Hamiltonian dynamics;
several of these have been obtained, in particular
a variational formulation and a study of (quaternionic) integrable
systems \cite{GM,GMspt,GM2,MT}.

A key ingredient which is obviously missing from this parallel is that
of {\it canonical} transformations; it should be stressed that this is
of independent interest: characterizing the group of transformations
which leave a given hyperkahler structure invariant (in a sense to be
detailed below) is of interest independently of the hyperhamiltonian
dynamics motivation (indeed in this sense it has been studied in the
literature \cite{Sa89}; thus, albeit our approach is from the point of view of hyperhamiltonian dynamics, it is not surprising that some of our findings will reproduce -- with different approach and methods -- results which are known from Differential Geometry \cite{Sa89}).

The purpose of this work is to start a detailed study of this problem taking into account previous results of the theory of hyperkahler manifolds and introducing new concepts and tools to fit them into the frame of hyper-hamiltonian dynamics, that is, to
properly define -- and then study -- canonical transformations for
hyperhamiltonian dynamics and hyperkahler structures.
As well known, in the symplectic (or Hamiltonian) case canonical
transformations can be defined in several equivalent ways (see
e.g. \cite{Arn}); direct naive extensions of these to the
hyperkahler framework are equally not viable, so suitable
generalizations should be considered, and it turns out
generalizations starting from notions which are equivalent in the
symplectic case will produce non-equivalent notions in the
hyperkahler case.

We will thus distinguish between \hK (or equivalently
hypersymplectic) transformations, preserving in a certain sense
the hyperkahler structure (and the associated hypersymplectic one),
see Definition \ref{sec:maps}.2, and canonical transformations,
preserving a certain four forms associated to the \hK structure,
see Definition \ref{sec:maps}.3 and more generally the discussion
of Sect.\ref{sec:maps}. In this sense, it is not entirely trivial
that our discussion, based on a dynamical systems point of view, ends
up on the one hand focusing on concepts already used in the differential
geometric approach \cite{Sa89, Jo00}, but also, on the other hand,
showing that preservation (in a suitable sense) of the hyper-Kahler
structure is not the natural requirement to be considered dynamically --
we will in fact  distinguish \hK and canonical transformations, see
below.

The present paper focuses to a large extent on the discussion of
what are the suitable generalizations mentioned above, i.e. what
are the appropriate definitions of \hK and canonical
transformations in general (see Sect.\ref{sec:maps});
we will also characterize them by providing
equations to be satisfied by the transformations.
In a companion paper \cite{GReuclidean} we will obtain a full
characterization (that is, we solve the characterizing equations) of \hK maps in the
Euclidean case; this is related to the general case via the result
presented here in Sect.\ref{sec:hkmaps}. Albeit such a full characterization
is obtained only in the Euclidean (flat) case, it should be emphasized
that this covers a number of physically relevant cases \cite{GR}:
not only  the Dirac equation -- which can be recast in terms of
hyperhamiltonian  dynamics -- lives in flat Minkowski space, but many of
the physically relevant non-flat \hK manifolds and structures are
obtained through a momentum map type construction \cite{HKLR} from
Euclidean $\R^{4 n}$ with standard \hK structures (see e.g.
\cite{GRtnut}.

As for canonical transformations, these are characterized in  Sect.\ref{sec:canmaps}. The key requirement, based on a reformulation of the familiar area condition for canonical transformations in symplectic dynamics, will be the preservation of a certain four-form, invariantly attached to the quaternionic structure identified by the \hK one. Application to the Dirac equation requires to consider dual \hK structures, as discussed in \cite{GR}; it is thus natural to consider and study \hK and canonical maps for these as well, which is  done in Sect.\ref{sec:dual}.

We will then finally consider our original motivation, namely  hyperhamiltonian  dynamics (Sect.\ref{sec:hhcan}). It will turn out that  this does not necessarily preserve the \hK structure (in any of the senses discussed in previous sections). Our main result  in this context  (holding in general, i.e. with no limitation to the Euclidean case) will be  that, similarly to what happens for Hamilton dynamics, {\it a  hyperhamiltonian flow generates a one-parameter group of canonical  transformations}.

\bigskip\noindent
{\it Notational convention.} We will consider smooth real
manifolds $M$ of dimension $4 n$, equipped with a Riemannian
metric and three structures of several types (complex, Kahler,
symplectic). We will be using Latin indices (running from $1$ to
$m = 4n$) for the local coordinates on the manifold $M$, and Greek
letters for the label (running from 1 to 3) attached to the
complex (Kahler, symplectic) structures on $M$; note that we
should distinguish between covariant and contravariant Latin
indices, as we deal with a generic Riemannian metric $g$, while
the metric in the $\a$ space is Euclidean, i.e. Greek labels could
be written equally as lower or upper indices (and we will
sometimes move them for typographical convenience). The Einstein
summation convention will be used unless otherwise stated; when
confusion could arise we will indicate explicitly summation.

\section{Hyper-Kahler structures}
\label{sec:bkI}

Let us start by recalling some basic definitions, mainly of
geometrical nature, which we will use in the following (see e.g.
\cite{AlM,At2,AtH} for further detail). All manifolds and
related geometric objects to be considered will always be real and
smooth; we will sometimes omit to indicate this for the sake of
brevity.

Let $(M,g)$ be a smooth real Riemannian manifolds; as well known
there is a unique torsion-free metric connection on it, the
Levi-Civita connection $\grad$.

\subsection{Kahler manifolds}

Consider a smooth real Riemannian manifold $(M,g)$ of dimension
$m=2k$. An almost complex structure on this is a field of
orthogonal transformations in $\T M$, i.e. a (1,1) type tensor
field $J$ such that $J^2 = - I$, with $I$ the identity map.

A {\it Kahler manifold } $(M,g,J)$ is a smooth orientable real
Riemannian manifold $(M,g)$ of dimension $m=2k$ equipped with an
almost-complex structure $J$ which has vanishing covariant
derivative under the Levi-Civita connection, $\nabla J = 0$.

Note the latter condition actually implies -- due to the
Newlander-Nirenberg theorem \cite{NN} -- the integrability of
$J$; so $(M,g,J)$ is a complex manifold.

The two-form $\om \in \La^2 (M)$ associated to $J$ and $g$ via the
Kahler relation \beq\label{eq:kahler} \om (v,w) \ = \ g (v,Jw)
\eeq is closed and non-degenerate; hence it defines a symplectic
structure in $M$, and each Kahler manifold is also symplectic.
(The converse is not true, and there are symplectic manifolds
which do not admit any Kahler structure.)

\subsection{Hyperkahler manifolds}

A {\it hyperkahler manifold } is a real smooth orientable
Riemannian manifold $(M,g)$ of dimension $m = 4n$ equipped with
three almost-complex structures $J_1,J_2,J_3$ which:
\begin{itemize} \item[(i)] are covariantly constant under the Levi-Civita
connection, $\nabla J_\a = 0$ (hence they are actually complex
structures on $(M,g)$, see above); and \item[(ii)] satisfy the
quaternionic relations, i.e. \beq\label{eq:quaternionic} J_\a \,
J_\b \ = \ \epsilon_{\a \b \ga} \, J_\ga \ - \ \delta_{\a \b} \, I
\eeq with $\epsilon_{\a \b \ga}$ the completely antisymmetric
(Levi-Civita) tensor.
\end{itemize}

Simple examples of hyperkahler manifolds are provided by
quaternionic vector spaces ${\bf H}^k$ and by the cotangent bundle
of complex manifolds \cite{Jo00}.

Note that the relations \eqref{eq:quaternionic} imply that the
$J_\a$ satisfy the $SU(2)$ commutation relations, but also involve
the multiplication structure.

We denote the ordered triple ${\Jb} = (J_1 , J_2 , J_3)$ as a {\it
hyperkahler structure } on $(M,g)$. We will denote a hyperkahler
manifold as $(M,g;J_1,J_2,J_3)$, or simply as $(M,g;\Jb)$.

Obviously a hyperkahler manifold is also Kahler with respect to
any linear combination $J = \sum_\a c_\a J_\a$ such that $|c|^2 :=
c_1^2 + c_2^2 + c_3^2 = 1$; thus we have a $S^2$ sphere of Kahler
structures on $M$. More precisely, we introduce the space
\beq\label{eq:quatstruct} {\bf Q} \ := \ \big\{ \sum_\a c_\a J_\a
\ , \ c_\a \in {\bf R} \big\} \ \approx \ {\bf R}^3 \ , \eeq also
called the {\bf quaternionic structure} on $(M,g)$ spanned by
$(J_1,J_2,J_3)$ \cite{AlM}; and denote by ${\bf S} \approx S^2$
the unit sphere in this space. Points in ${\bf S}$ are in one to
one correspondence with those Kahler structures on $(M,g)$ which
are in the linear span of the given basis structures $J_\a$, and
opposite points correspond to complex conjugate structures. The
sphere ${\bf S}$ will play a central role in our discussion and
deserves a special name.

\medskip\noindent
{\bf Definition \ref{sec:bkI}.1.} {\it The unit sphere in ${\bf
Q}$, i.e. the set \beq {\bf S} \ := \ \big\{ \sum_\a c_\a J_\a \ ,
\ c_\a \in {\bf R} \ : \ |c|^2 := \sum_\a c_\a^2 = 1 \big\} \
\approx \ S^2 \ss {\bf R}^3 \ , \eeq is the {\bf Kahler sphere}
corresponding to the hyperkahler structure $\Jb = (J_1,J_2,J_3)$.}

\medskip\noindent
{\bf Definition \ref{sec:bkI}.2.} {\it Two hyperkahler structures
on $(M,g)$ defining the same quaternionic structure ${\bf Q}$, and
hence the same Kahler sphere ${\bf S}$, are said to be {\bf
equivalent}. An equivalence class of hyperkahler structures is identified
with the corresponding quaternionic structure,  and viceversa.}

\medskip\noindent
{\bf Remark \ref{sec:bkI}.1.} Note that the quaternions ${\bf H}$
act (linearly) in a natural way on ${\bf Q}$; moreover, the group
${\bf H}_0 $ of quaternions of unit norm acts preserving ${\bf
S}$. More precisely, unit quaternions other than the real unit
(which acts by the identity) will generate a rotation of the
sphere ${\bf S} \simeq S^2$. The action mentioned here is of
course given, for the quaternion $h = h_0 + i h_1 + j h_2 + k
h_3$, by $J \mapsto H^{-1} J H$, with $H = h_0 I + h_1 J_1 + h_2
J_2 + h_3 J_3$ (where $h_i \in \R$). In the case $|h|=1$, where of
course $|h| := (h_0^2 + h_1^2 + h_2^2 + h_3^2)^{1/2}$, we have
$H^{-1} = h_0 I - h_1 J_1 - h_2 J_2 - h_3 J_3$. \EOR

\medskip\noindent
{\bf Remark \ref{sec:bkI}.2.} There are obvious symplectic
counterparts to the notions defined above, the correspondence
being through the Kahler relation \eqref{eq:kahler} (note this
implies that the metric will play a role, at difference with the
standard symplectic case; in fact, in order to have a more usual
analogue one should think of the Kahler case). Thus the symplectic
forms $\om_\a$ correspond to the $J_\a$, and
$(M,g;\om_1,\om_2,\om_3)$ is a {\it hypersymplectic manifold}. Any
nonzero linear combination of the $\om_\a$, i.e. any $\mu \not= 0$
in \beq {\mathcal Q} \ := \ \big\{ \mu = \sum_\a c_\a \om_\a \ , \
c_\a \in {\bf R} \big\} \ \approx \ {\bf R}^3 \eeq is also a
symplectic structure on $M$; in other words we have a punctured
three dimensional space ${\bf R}^3 \backslash \{ 0 \}$ of
symplectic structures in $M$. Denote by ${\mathcal S}$ the unit
sphere in ${\mathcal Q}$; the $\mu \in {\mathcal S}$ are {\it
unimodular} symplectic structures in $M$. Obviously the sphere
${\mathcal S}$ corresponds to ${\bf S}$ via the Kahler relation;
hence ${\mathcal S}$ is the {\bf symplectic Kahler sphere} for the
hyperkahler structure $(J_1,J_2,J_3)$, and two hypersymplectic
structures defining the same ${\mathcal S}$ are {\it equivalent}.
\EOR

\subsection{Relations between equivalent structures}

The notion of equivalent structures will play a key role in a
large part of the following; it is thus worth presenting some
remarks to further characterize them.

Let us consider two equivalent structures $\Jb$ and $\wt{\Jb}$; by
definition these generate the same three-dimensional linear space
${\bf Q}$, hence each of them can be written in term of the other.
In particular we can write
\beq\ \wt{J}_\a \ = \ R_{\a \b} \, J_\b \eeq
(the metric in ${\bf Q}$ is Euclidean, so we will write both
indices as lower ones for typographical convenience). Now the
requirement that $\wt{J}_\a^2 = J_\a^2 = - I$ forces the (real,
three-dimensional) matrix $R$ to be orthogonal, $R \in O(3)$.
Moreover, the quaternionic relations \eqref{eq:quaternionic} require
$\Det (\wt{J}_\a) = \Det (J_\a) = 1$, hence also
$\Det (R) = 1$; in other words we must actually have $R \in SO(3)$.

The same argument also applies to equivalent \hS structures: in
this case we also conclude that if $\{ \om_1 , \om_2 , \om_3 \}$
and $\{ \wt{\om}_1 , \wt{\om}_2 , \wt{\om}_3 \}$ are equivalent
\hS structures, then necessarily $\wt{\om}_\a = R_{\a \b} \om_\b$,
with $R \in SO(3)$.

\medskip\noindent
{\bf Remark \ref{sec:bkI}.4.} A simple example of $R \in O(3)$
but not in $SO(3)$, for which it is immediate to check that the
Kahler sphere is mapped into itself but the quaternionic relations
are not preserved, is given by $R = {\rm diag} (-1,1,1)$. \EOR

\medskip\noindent
{\bf Remark \ref{sec:bkI}.5.} Here we are considering generic maps
in ${\bf Q}$ or ${\bf S}$ (or more generally in the space of
tensors defined on $M$). If we consider only maps induced by maps
in $M$, then the situation is different. In particular, due to
their tensorial nature, the quaternionic relations are
automatically preserved under any (non-singular) map $\phi : M \to
M$. \EOR

\subsection{Standard structures in $\R^4$}
\label{sec:stand_sub}

In the following we will make reference to ``standard'' \hK and
\hS structures in $\R^{4n}$; these are obtained from standard
structures in $\R^4$ (with Euclidean metric) \cite{GM,GR}. We will
consider the standard volume form $\Om = \d x^1 \w \d x^2 \w \d
x^3 \w \d x^4$ in $\R^4$.

There are two such standard structures, differing for their
orientation. The positively-oriented standard \hK structure is
given by {\small \beq\label{eq:standhKpos}    \mathcal{Y}_1 =
\pmatrix{0&1&0&0\cr -1&0&0&0\cr 0&0&0&1\cr 0&0&-1&0\cr}, \quad \mathcal{Y}_2
= \pmatrix{0&0&0&1\cr 0&0&1&0\cr 0&-1&0&0\cr -1&0&0&0\cr},\quad
\mathcal{Y}_3 = \pmatrix{0&0&1&0\cr 0&0&0&-1\cr -1&0&0&0\cr 0&1&0&0\cr} \ .
\eeq } To these complex structures correspond the symplectic
structures, satisfying $(1/2) (\om_\a \w \om_\a) = \Om$ (no sum on
$\a$), \beq\label{eq:standhSpos}
 \begin{array}{ll}
\om_1 \ = \ \d x^1 \w \d x^2 \, + \, \d x^3 \w \d x^4 \ , &
\om_2 \ = \ \d x^1 \w \d x^4 \, + \, \d x^2 \w \d x^3 \ , \\
\om_3 \ = \ \d x^1 \w \d x^3 \, + \, \d x^4 \w \d x^2 \ . &
\end{array} \eeq

The negatively-oriented standard \hK structure is given by
{\small \beq\label{eq:standhKneg}
\hat\mathcal{Y}_1 = \pmatrix{0&0&1&0\cr 0&0&0&1\cr
-1&0&0&0\cr 0&-1&0&0\cr}, \quad \hat\mathcal{Y}_2 = \pmatrix{0&0&0&-1\cr
0&0&1&0\cr 0&-1&0&0\cr 1&0&0&0\cr}, \quad \hat\mathcal{Y}_3 =
\pmatrix{0&-1&0&0\cr 1&0&0&0\cr 0&0&0&1\cr 0&0&-1&0\cr } \eeq }
In this case, to these complex structure correspond the symplectic structures
\beq\label{eq:standhSneg}
 \begin{array}{ll}
\^\om_1 \ = \ \d x^1 \w \d x^3 + \d x^2 \w \d x^4 \ , &
\^\om_2 \ = \ \d x^4 \w \d x^1 + \d x^2 \w \d x^3 \ , \\
\^\om_3 \ = \ \d x^2 \w \d x^1 + \d x^3 \w \d x^4 \ ; &
\end{array} \eeq
these satisfy $(1/2) (\om_\a \w \om_\a) = - \Om$ (again with no
sum on $\a$).

\medskip\noindent
{\bf Remark \ref{sec:bkI}.6.} Note that $[Y_\a , \^Y_\b ] = 0$ for
all $\a , \b$. The existence of these two equivalent (and
oppositely oriented) mutually commuting real representations of
$su(2)$ (and hence of the group $SU(2)$ as well) is of course
related to the quaternionic nature of $SU(2)$ in the
classification given by the real version of Schur Lemma (see e.g.
chapt.8 of \cite{Kir}, in particular thm.3  there).  \EOR

\medskip\noindent
{\bf Remark \ref{sec:bkI}.7.} Note also that while the $SU(2)$
commutation relations are satisfied by any representation, the
condition $J_\a^2 = - I$ imply that the tensors $J_\a$ are
represented, at any given point, by a sum of copies of the two
(oppositely oriented) fundamental representations, i.e. the
standard ones defined above. \EOR

\medskip\noindent
{\bf Remark \ref{sec:bkI}.8.} The orientation of hyperkahler
structures is detected by an algebraic invariant (of matrices
representing the complex structures $J_\a$), defined on
generic matrices $A$ of order $2 m$ as \beq \mathcal{P}_m (A) \ :=
(1/p_m) \ \sum_{i_s,j_s=1}^{2 m} \ \epsilon_{i_1 j_1 ... i_m j_m}
\ A_{i_1 j_1} ... A_{i_m j_m} \ , \eeq with $p_m = 2^m (m!)$ a
combinatorial coefficient. This will appear in Section
\ref{sec:canmaps} and is discussed in Appendix A. It is immediate
to check that $\mathcal{P}_2 (\mathcal{Y}_\a) = 1$, $\mathcal{P}_2 ({\hat\mathcal{Y}}_\a
) = -1 $. \EOR

\subsection{HyperKahler structures in coordinates}

The results we want to prove are of local nature, so we
can work on a single chart of the hyperkahler manifold
$(M,g;\Jb)$. In the following we will use local coordinates $x^i$
($i = 1, ... , 4n$); it will be useful to have a standard notation
for expressing the objects introduced above in coordinates.

The metric $g$ is defined in coordinates by $g_{ij} \, \d
x^i \, \d x^j$ (we will use the same letter for its corresponding matrix); when using shorthand notation (with no
indices) we will denote the contravariant metric tensor $g^{ij}$
by $g^{-1}$.

The complex structures $J_\a$ and the associated Kahler symplectic
forms $\om_\a$ will be written as \beq \begin{array}{l}
J_\a \ = \ {(Y_\a)^i}_j \ \pa_i \otimes \d x^j \\
\om_\a \ = \ \ (K_\a)_{ij} \ \d x^i \w \d x^j \ ;
\end{array} \eeq
where the wedge product is defined as $\d x^i \wedge \d x^j =(1/2)(\d x^i\otimes \d x^j-\d x^j\otimes \d x^i)$. We will also consider tensors of type $(2,0)$ associated to these,
i.e. \beq M_\a^{ij} \ = \ g^{i\ell} \, K^\a_{\ell m} \, g^{mj} \ .
\eeq

Note that here $M_\a, Y_\a , K_\a$ are in general functions of the
point $x$, and are of course not independent (we prefer to
have distinct notations for the tensor fields $Y_\a , K_\a = g
Y_\a , M_\a = Y_\a g^{-1}$ as these will be useful in writing
subsequent equations in compact form without the need to write
down all the indexes; note $K_\a^{-1} = - M_\a$, and of course $Y_\a^{-1} = - Y_\a$).

The quaternionic relations \eqref{eq:quaternionic} are reflected
into the same relations being satisfied by the matrices $Y_\a$,
and similar ones -- involving also $g$ -- by the $K_\a$ and
$M_\a$, i.e. \beq\label{eq:quatrelcomp} \begin{array}{ll}
Y_\a \, Y_\b \ &= \ \epsilon_{\a\b\gamma} \, Y_\gamma \ - \ \delta_{\a\b} \, I \\
K_\a \, g^{-1} \, K_\b \ &= \ \epsilon_{\a\b\gamma} \, K_\gamma \
- \ \delta_{\a\b} \, g \\
M_\a \, g \, M_\b \ &= \ \epsilon_{\a\b\gamma} \, M_\gamma \ - \
\delta_{\a\b} \, g^{-1} \ . \end{array} \eeq

Similarly, the fact that the $J_\a$ are covariantly constant
implies that $\grad Y_\a = 0$ as well; as $g$ is by definition
also covariantly constant under its associated Levi-Civita
connection, we also have $\grad K_\a = 0$, $\grad M_\a = 0$.

\section{Hyperkahler and canonical transformations}
\label{sec:maps}

In this Section we will set our definitions of hyperkahler
(or, in its case, quaternionic, see below)
and of canonical transformations. These will be built by (non-trivial)
analogy with the standard case of canonical transformations in
Hamiltonian mechanics. We will thus start by briefly recalling
this standard case, referring e.g. to \cite{Arn} for details.

\subsection{Symplectic maps}

Let $(M,\om)$ be a symplectic manifold (of dimension $2n$); we say
that a map $\phi : M \to M$ is {\it symplectic} if it preserves
the symplectic form $\om$, i.e. if \beq\label{eq:symptrans} \phi^*
(\om ) \ = \ \om \ . \eeq

An equivalent characterization is also quite common. As well
known, by Darboux theorem \cite{Arn} one can introduce local
coordinates $(p_a,q^a)$, $a=1,...,n$, in a neighborhood $U \ss M$
such that $\om = \d p_a \w \d q^a$. Then, one considers local
minimal manifolds on which $\om$ is non-degenerate; these are
two-dimensional and are spanned by $q^a$ and $p_a$ (with same
$a$). They are known as Darboux submanifolds and denoted as $U_a$.
Denote by $\iota_a$ the embedding $\iota_a : U_a \hookrightarrow U
\sse M$; then the restriction $\iota_a^* \om$ of the symplectic
form to $U_a$ provides a volume form $\Om_a = \d p_a \w \d q^a$
(no sum on $a$) on $U_a$. Then, for any two-chain $A$ in $U$ and
with $\pi_a : U \to U_a$ the projection to $U_a$,
$$ \int_A \om \, = \, \int_A \sum_{a=1}^n \d p_a \w \d q^a \, = \,
\sum_{a=1}^n \int_A \Om_a \ = \ \sum_{a=1}^n \mathrm{area} [\pi_a A] \ ; $$
thus preservation of $\om$ is equivalent to preservation of the
sum of oriented areas of projection of any $A$ to Darboux
submanifolds. That is, the map $\phi$ is canonical if
$$ \sum_{a=1}^n \mathrm{area} [\pi_a A] \ = \
\sum_{a=1}^n \mathrm{area} [\pi_a (\phi A)] \ . $$

It should be noted that if we start from a manifold equipped with
a Riemannian metric, passing to Darboux coordinates will in
general not preserve it. Thus this construction is is general not
viable if one requires preservation of the metric.

In the case of a Kahler manifold, the symplectic form $\om$
corresponds to a complex structure $J$ through the Kahler relation
\eqref{eq:kahler}. This satisfies $J^2 = - I$, and provides a
splitting of $\T_0 M$ (at any point $m_0 \in M$) into
two-dimensional invariant subspaces; the volume form $\Om$ defined
in $M$ induces volume forms $\Om_a$ in each of these, and $\om =
\sum \Om_a$. Thus again canonical transformations can be
characterized as those satisfying \beq\label{eq:cantrans}
\sum_{a=1}^n \Om_a \ = \ \sum_{a=1}^n \phi^* (\Om_a ) \ . \eeq

Note this construction does not make use of Darboux coordinates or
manifolds, but only of the splitting of $\T M$ induced by the
action of the complex structure; moreover, we only consider volume forms.

\subsection{Hyperkahler transformations}

Let us now pass to consider hyperkahler structures. As already
noted, the tensorial nature of the quaternionic relation
\eqref{eq:quaternionic} guarantees they will be preserved under
any map $\phi : M \to M$. Note also that here the Riemannian
metric is an essential part of the structure, so if we look at
maps which preserve the hyperkahler (or the quaternionic)
structure it is natural to only consider maps $\phi : M \to M$
which are orthogonal with respect to $g$.

\subsubsection{Strongly hyperkahler maps}

It may seem natural to generalize \eqref{eq:symptrans} by
requiring that the three symplectic forms $\om_\a$ (and hence all
symplectic form in $\mathcal{Q}$) are  preserved; from the point
of view of the complex structures, this means considering {\it
tri-holomorphic} maps. However, this criterion would be
exceedingly restrictive, and we will deal with a weaker form of
it. We will reserve a different name for this case.

\medskip\noindent
{\bf Definition \ref{sec:maps}.1.} {\it Let $(M,
g;J_1,J_2,J_3)$ be a hyperkahler manifold. We say that
the orthogonal map $\phi : M \to M$ is {\bf strongly
hyperkahler} if it leaves the three complex structures $J_\a$
invariant.}
\bigskip

\medskip\noindent
{\bf Remark \ref{sec:maps}.1.} We have stated that this class of maps
is exceedingly restrictive. To show this is the case, let us consider
the map generated by a Hamiltonian flow, say under the symplectic
structure $\om_1$. It is easy to check that in this case (even in
the simplest integrable case, with hamiltonian $|x|^2 / 2$), the forms
$\om_2 , \om_3$ are not preserved. In fact for the transformed forms
$\wt{\om}_\a$ we have
$$ \wt{\om}_1 = \om_1 , \ \wt{\om}_2 = \cos(\th ) \om_2 - \sin (\th ) \om_3 ,
\  \wt{\om}_3 = \sin(\th ) \om_2 + \cos (\th ) \om_3 \ ; $$ here
$\th$ is an angle, depending on time. Thus the forms $\om_2 ,
\om_3$ are rotated in the plane they span in $\Q$. In other words,
the hypersymplectic structure is in this case mapped into an
equivalent -- but different -- one. \EOR

\subsubsection{Hyperkahler maps}

The above remark suggest that (as discussed also in
\cite{GM,GMspt}) the appropriate generalization of symplectic
transformations in the hyperkahler case should {\it not } require
the preservation of the three symplectic (Kahler) forms; we should
rather require -- beside the preservation of the metric -- the
milder condition that the hyperkahler structure is mapped into an
equivalent one.

\medskip\noindent
{\bf Definition \ref{sec:maps}.2.} {\it Let $(M,
g;\Jb)$ be a hyperkahler manifold. We say that
the orthogonal map $\phi : M \to M$ is {\bf hyperkahler} if it
maps the hyperkahler structure into an equivalent one, i.e. if
$\phi^* : {\bf S} \to {\bf S}$.}

\medskip\noindent
{\bf Remark \ref{sec:maps}.2.} With this definition, the
Hamiltonian flow considered in Remark \ref{sec:maps}.1 will
generate a one-parameter group of hyperkahler maps. Note that a
generic Hamiltonian flow will not preserve the metric and hence
will not qualify as generating (a family of) hyperkahler maps.
\EOR

\medskip\noindent
{\bf Remark \ref{sec:maps}.3.} Hyperkahler maps will preserve the
quaternionic structure; we will thus also refer to them as {\bf
quaternionic} maps. \EOR
\bigskip

Finally, we note that, as obvious, the concepts considered in this
section can also be expressed referring to symplectic (rather than complex) structures; we will in this framework have the corresponding

\medskip\noindent
{\bf Definition \ref{sec:maps}.1'.} {\it Let $(M,
g;\om_1,\om_2,\om_3)$ be a hypersymplectic manifold. We say that
the orthogonal map $\phi : M \to M$ is {\bf strongly
hypersymplectic} if it leaves the hypersymplectic structures
invariant, i.e. if $\phi^* (\om_\a) = \om_\a$ for $\a=1,2,3$.}

\medskip\noindent
{\bf Definition \ref{sec:maps}.2'.} {\it If $(M, g;\om_1,\om_2,\om_3)$
is a hypersymplectic manifold, we say that the orthogonal map
$\phi : M \to M$ is {\bf hypersymplectic} if it maps the
hypersymplectic structure into an equivalent one, i.e. if $\phi^*
: {\mathcal S} \to {\mathcal S}$.}

\subsection{Canonical transformations}

We will reserve the name ``canonical transformations'' (or maps)
for those which satisfy the (generalization of) the criterion
based on conservation of projected areas, see \eqref{eq:cantrans}.

In the hyperkahler case, the three complex structures induce a
splitting of $\T_0 M$ (at any given point $m_0 \in M$ of the
$4n$-dimensional manifold $M$) into four-dimensional invariant
subspaces $U_a$. (It may be worth remarking again, in this
respect, that the quaternionic relations \eqref{eq:quaternionic}
imply the $J_\a$ satisfy the $su(2)$ Lie algebra commutation
relations, but also involve the multiplicative structure. In
particular, they imply that the $J_\a$ (at a given point) provide
a representation of $su(2)$ as the sum of $n$ four-dimensional
real irreducible representations.)

\subsubsection{The Euclidean case}

In the Euclidean case (thus $M = \R^{4 n}$ and $g = I_{4n}$) the
Levi-Civita connection is flat and the $J_\a$ are actually
constant (it is easy to see that in this case the complex
structures are given by a sum of structures in standard form).
Thus the splitting actually applies to the full $M$ (the
decomposition is of course in terms of $\R^4$ subspaces). More
generally, if $M$ is locally Euclidean,  the invariant
four-dimensional subspaces of $\T_x M$ (for $x \in U \ss M$) form
a distribution which has invariant four-dimensional integral
manifolds $U_a$. Considering the embedding $\iota_a : U_a
\hookrightarrow U$, the volume form on $U_a$ is obtained as \beq
\Om_a \ = \ \iota_a^* \( \frac{1}{2} \ \om \w \om \) \ , \eeq for
$\om$ any symplectic form in $\mathcal{S}$. (That this is
independent of $\om \in \mathcal{S}$ is easily checked via the
explicit form of the $K_\a$ in standard form, and the remark made
above that in the Euclidean case the structures can be written in
standard form of either orientation. Actually, this remark
amounts, in Lie theoretic terms, to the fact that there are only
two real irreducible representations of the $su(2)$ Lie algebra of
dimension four.)

It follows easily that the maps which preserve the sum of oriented
volumes, thus the sum of the $\Om_a$, are precisely those which
preserve the four-forms $\om \w \om$ for any $\om \in
\mathcal{S}$, and in particular for $\om = \om_\a$ (with $\a =
1,2,3$).

\subsubsection{The general case}

Motivated by the above discussion for the Euclidean case, we will
extend the characterization of canonical transformations found in
that case to the general situation.

\medskip\noindent
{\bf Definition \ref{sec:maps}.3.} {\it Let $(M, g;\Jb)$ be a
hyperkahler manifold, and $\mathcal{Q}$ the corresponding
symplectic Kahler sphere. We say that the map $\phi : M \to M$ is
{\bf canonical} if, for any $\om \in \mathcal{S}$, it preserves
the form $ \om \w \om$.}
\bigskip

\bigskip\noindent
{\bf Remark \ref{sec:maps}.4.} It is clear that the two
notions of canonical and hyperkahler (or quaternionic) maps
proposed here are {\it not } equivalent (at difference with the
notion holding in the symplectic or Kahler case which they
generalize). In a way, quaternionic maps preserve the quaternionic
structure, while canonical ones only preserve the square of forms
associated to it; moreover, note that we are {\it not} requiring
canonical maps to be orthogonal. Consider e.g. $\om_1$ (see
Section \ref{sec:bkI}): under the map $x^1 \to \la x^1$, $x^2 \to
\la x^2$, $x^3 \to \la^{-1} x^3$, $x^4 \to \la^{-1}  x^4$, the
form $\om_1$ is not preserved (note $g$ is not preserved as well)
nor mapped to a different form in $\mathcal{S}$, but $\om_1 \w
\om_1$ is invariant. More generally, a  canonical map could even
mix the positively and negatively oriented structures. \EOR
\bigskip

The criterion for a transformation to be canonical can also be
stated in terms of a basis for $\mathcal{S}$, i.e. of the $\om_\a$
associated to the $J_\a$. In terms of these we have the equivalent
definition:

\medskip\noindent
{\bf Definition \ref{sec:maps}.4.} {\it The map $\phi : M \to M$
is {\bf canonical} for the hyperkahler structure $(g;\Jb)$ if and
only if (with no sum on $\a$)
$$ \iota_a^* (\om_\a \w \om_\a ) \ = \
\iota_a^* [ \phi^* (\om_\a \w \om_\a ) ] \ \ \ \a = 1,2,3 \ , \ \ a = 1,...,n \ . $$}

\medskip\noindent
{\bf Remark \ref{sec:maps}.5.} In order to see that this is
equivalent to the previous one, it suffices to note that any $\mu
\in {\mathcal S}$ is written as $\mu = c_\a \om_\a$, and that
independence of the $\om_\a$ (required by the quaternionic
relations) imply that $\iota_a^* (\om_\a \w \om_\b ) = 0$ when $\a
\not= \b$. Thus
$$ \iota_a^* (\mu \w \mu ) \ = \ \sum_{\a,\b = 1}^3 \, c_\a c_\b \,
\iota_a^* (\om_\a \w \om_\b) \ = \ \sum_{\a=1}^3 \, c_\a^2 \iota_a^*
(\om_\a \w \om_\a ) \ . $$ Given the arbitrariness of $\mu$, i.e.
of the $c_\a$, we conclude that indeed Definition \ref{sec:maps}.4
is equivalent to Definition \ref{sec:maps}.3. \EOR

\section{Characterization of hyperkahler maps}
\label{sec:hkmaps}

We can now discuss hyperkahler (quaternionic) transformations,
i.e. applications $\Phi : M \to M$ which map the hyperkahler
structure into an equivalent one. It is clear that these form a
group, which will be denoted as $\hSp$, or more precisely $\hSp
(M, g,{\bf J})$.
It is obvious that (see Definition \ref{sec:maps}.2') this is
equivalently the group of hypersymplectic transformations.
(In fact, the notation $\hSp$ stands for ``hypersymplectic''.)

{It should be stressed that there is an essential difference between this and the symplectic group which is familiar from standard Hamiltonian dynamics or from symplectic geometry. In fact, in the symplectic case the Darboux theorem allows to reduce (locally) any symplectic structure to the standard form $\om = \d p_i \w \d q^i$; with this $\om$ is (locally) constant, and the maps which preserve $\om$ at a given point $x_0$ will also -- when extended as constant ones -- preserve it in a full neighborhood of $x_0$. Thus, as well known, one effectively reduces to a problem in linear algebra. On the other hand, there is no Darboux theorem for \hK structures, and the latter are in general {\it not} constant (even locally), but instead covariantly constant. Thus the analysis made at a single reference point $x_0$ will not immediately provide ``hypersymplectic'' maps in a neighborhood of it, the extension requiring to have covariantly constant maps.}

\subsection{Hyperkahler maps for Euclidean versus general manifolds}

Characterization of hyperkahler maps is much easier in the
Euclidean case -- where we can in practice reduce to consider the
standard structures introduced in Section \ref{sec:bkI} -- than in
the general one, even at the local level. In both cases, one would
like first to reduce the structure to standard form at least in a
reference point (for the Euclidean case this will hold on the
whole manifold). Note that in the following we will sometime say, for ease of writing, ``new metric'' (and so on) to mean ``expression of the metric in the new
coordinates'' (and so on); we hope the reader will forgive this
little abuse of language.

Let us consider a neighborhood $U \ss M$ and local coordinates
$x^i$ in it; we denote the covariant derivative w.r.t. $x^i$
defined by the Levi-Civita connection as $\nabla_i$. This acts on
$(1,1)$ tensor fields $J$ as $\nabla_i J = \pa_i J + [A_i , J]$;
hence the $J_\a$ satisfy \beq \nabla_i \, J_\a \ = \ \pa_i \, J_\a
\ + \ [A_i , J_\a ] \ = \ 0 \ . \eeq

Let us now consider a change of variables; we denote its Jacobian
by $\La$, i.e. $\La^i_{\ j} = (\pa x^i / \pa \wt{x}^j)$. Under
this change of coordinates the metric, represented in the old
coordinates by the matrix $g$ is represented by the matrix
$\wt{g}$ with \beq \wt{g} \ = \ \La^T \, g \, \La \ ; \eeq
correspondingly the coordinate expression of the Levi-Civita
connection changes (we write $\wt{\nabla}$ for the new
expression), and the covariant derivatives under this (acting on
(1,1) tensor fields) are written in coordinates as \beq
\wt{\nabla}_i \ = \ \wt{\pa}_i \ + \ [\wt{A}_i , . ] \ , \eeq
where \beq \wt{A}_i \ = \ \La^{-1} \, A_i \, \La \ - \ (\pa_i
\La^{-1} ) \, \La \ \equiv \ \La^{-1} \, A_i \, \La \ + \ \La^{-1}
\, (\pa_i \La) \ . \eeq The (1,1) tensor fields $J_\a$ are changed
into new (1,1) tensor fields $\wt{J}_\a$ with \beq \wt{J}_\a \ = \
\La^{-1} \, J_\a \, \La \ . \eeq

Obviously (as the considered relation do not depend on
coordinates) the $\wt{J}_\a$ are still orthogonal and covariantly
constant, and satisfy the quaternionic relations; in other words,
they are again a hyperkahler structure.

Let us now fix a reference point $x_0 \in U$, and choose a first
change of variables (with Jacobian $\La^{(0)}$) so that at this
point the new metric is just given by $\wt{g} (x_0) = \delta $,
which is always possible choosing a suitable $\La^{(0)}$.

If we want to consider further transformations which do not alter
the metric at this reference point, we have to consider only
changes such that their Jacobian, denoted by $\La^{(1)}$,
satisfies $\La^{(1)}(x_0) = B \in O (4n)$. It is quite clear that
by a suitable choice of this $B$, hence of the overall change of
coordinates with Jacobian $\La = \La^{(1)} \La^{(0)}$, we can
obtain that the new complex structures $\wt{J}_\a$ satisfy
$\wt{J}_\a (x_0) = Y_\a$ with $Y_\a$ the ``standard'' complex
structures considered in \cite{GM,GR} and given in
sect.\ref{sec:stand_sub}.

We summarize our discussion in the form of a Lemma.

\medskip\noindent
{\bf Lemma \ref{sec:hkmaps}.1.} {\it Given a hyperkahler manifold
$(M,g;\Jb)$ and a point $x_0 \in M$, it is always possible to change local coordinates around $x_0$ so that the metric and the complex
structures are written as $\wt{g}$, $\wt{J}_\a$, with $\wt{g} (x_0) = \de$, $\wt{J}_\a (x_0) = Y_\a$.}

\medskip\noindent
{\bf Remark \ref{sec:hkmaps}.1.} The Lemma deals with a single
point $x_0$; but, we are of course interested not only in what
happens at $x_0$, but at least in an open neighborhood $U$ of it.
The form of the $\wt{J}_\a$ at other points of $U$
is rather general, and only subject to the condition of being
covariantly constant, $\wt{\nabla} \wt{J}_\a = 0$. Note that if
$\wt{\nabla}$ has a nontrivial holonomy, this does not uniquely
define the $\wt{J}_\a$.
\EOR

\medskip\noindent
{\bf Remark \ref{sec:hkmaps}.2.} The holonomy group $\H$ must be a subgroup of the invariance group for the (integrable) quaternionic structure, i.e. $\h \sse \hSp$. We can expect that, unless the hyperkahler structure has some special (invariance) property, the two will just coincide (in \cite{GReuclidean} we will find this is the case in Euclidean spaces; this fact should be seen as a check that our notion of hyperkahler maps is an appropriate one. \EOR
\bigskip

Let us now discuss the relation between the groups $\hSp
(M,g,\Jb)$ and $\hSp_0 (4n) := \hSp (\R^{4n},g_0, \Jb_0)$; here and in
the following we denote by $\hSp_0 (4n)$ the group of \hS
transformations for metric in Euclidean form $g_0 = \delta$ and standard
\hK structures $\Jb_0$ at a reference point $x_0$.

\medskip\noindent
{\bf Lemma \ref{sec:hkmaps}.2.} {\it Let $R(x)$ be a map taking
$(g,{\bf J})$ into standard form  $(g_0,{\bf J}_0)$ at the point
$x_0$; then \beq\label{eq:hSp}    \hSp (4n,g,\Jb) \ = \ R^{-1}
(x_0) \, \hSp (4n,g_0,\Jb_0 ) \, R(x_0) \ = \ R^{-1} (x_0) \,
\hSp_0 (4n) \, R(x_0) \ . \eeq}

\medskip\noindent
{\bf Proof.} This just follows from $R : g \to g_0$ and $R : \Jb
\to \Jb_0$. Note that $R$ is not uniquely defined, as any map $\^R
= S \cdot R$ with $S = S(x)$ such that $S(x_0) \in \hSp_0 (4n)$
will have the same effect, but this lack of uniqueness will not
affect \eqref{eq:hSp}. \EOP

\subsection{Characterization for structures in standard form}

Thanks to Lemma \ref{sec:hkmaps}.2, we can just focus on $\hSp_0
(4n)$, i.e. deal with metric and \hK structures which are in
standard form at an arbitrary reference point $x_0$. We will from
now on write $g$, $\nabla$, $J_\a$, to denote the metric, the
associated connection, and the complex structures in this case.

In view of Definition \ref{sec:maps}.1, we have to look for
changes of coordinates $\Phi$ with Jacobian $\La$ which preserve
$g$ (and hence $\nabla$ and its coefficients $A_i$) and which map
$\Jb$ into an equivalent $\wt{\Jb}$. In other words we have to
require that \beq\label{eq:mapequiv} \wt{J}_\a \ = \ R_{\a \b} \,
J_\b \ \ \ \mathrm{with} \ \ \ R \in SO(3)  \ . \eeq

\medskip\noindent
{\bf Remark \ref{sec:hkmaps}.3.} One could think of a
generalization of \eqref{eq:mapequiv} with $R$ a matrix field with
values in $SO(3)$ rather than a constant one; this is actually
forbidden by the condition $\nabla \wt{J}_\a = 0$, which implies
$R$ is constant. In fact, we have immediately $ \nabla_i \wt{J}_\a
= \pa_i \wt{J}_\a + [A_i , \wt{J}_\a ] = \pa_i (R_{\a \b} J_\b ) +
R_{\a \b} [A_i , J_\b ] = (\pa_i R_{\a \b}) J_\b + R_{\a \b}
(\nabla_i J_\b ) = (\pa_i R_{\a \b}) \, J_\b$. Hence $\nabla
\wt{J}_\a = 0 $ if and only if $(\pa_i R_{\a \b}) = 0$, i.e. if
and only if the $R_{\a \b}$ are constant. \EOR
\bigskip

In order to discuss \eqref{eq:mapequiv} we will suppose that $J_\a
(x_0)$ is represented by $Y_\a$ in blocks $1,...,m$ and by
$\^Y_\a$ in blocks $m+1,...n$ (a reordering of blocks is needed
for the general case, but inessential).

It will be convenient to write the matrix $\La$ in terms of
four-dimensional blocks; we set a standard notation for this, and
write (no confusion should be possible between the sub-matrices
$A_{ij}$ and the connection coefficients $A_i$)
\beq\label{eq:lambda} \La \ = \ \pmatrix{ A_{11} & A_{12} & ... &
A_{1n} \cr A_{21} & A_{22} & ... & A_{2n} \cr ... & ... & ... &
... \cr A_{n1} & A_{n2} & ... & A_{nn} \cr} \ ; \ \ \La^T \ = \
\pmatrix{ A_{11}^T& A_{21}^T & ... & A_{n1}^T \cr A_{12}^T &
A_{22}^T & ... & A_{n2}^T \cr ... & ... & ... & ... \cr A_{1n}^T &
A_{2n}^T & ... & A_{nn}^T \cr} \ . \eeq

It will also be convenient to deal with $K_\a$ (rather than
$Y_\a$), so to avoid inversion of the matrix $\La$; the condition
$\wt{J}_\a = R_{\a \b} J_\b$ is equivalent to $\wt{K}_\a = R_{\a
\b} K_\b$.

We will write the (block-diagonal, once we pass to standard form)
matrices $K_\a$ and the (in general, not block-diagonal)
$\wt{K}_\a = \La^T K_\a \La$ as
$$ K^\a \ = \ \pmatrix{K^\a_{11} & 0 & ... & 0 \cr 0 & K^\a_{22} & ... & 0 \cr 0 & 0 & \ddots & 0 \cr 0 & ... & 0 & K^\a_{nn} \cr} \ , \ \ \wt{K}_\a \ = \ \La^T \, K_\a \, \La \ = \ \pmatrix{\wt{K}^\a_{11} & \wt{K}^\a_{12} & ... & \wt{K}^\a_{1n} \cr \wt{K}^\a_{21} & \wt{K}^\a_{22} & ... & \wt{K}^\a_{2n} \cr ... & ... & ... & ... \cr \wt{K}^\a_{n1} & ... & ... & \wt{K}^\a_{nn} \cr} \ . $$
It is easily checked that $\wt{K}^\a_{ij} = A_{\ell i}^T
K^\a_{\ell m} A_{mj}$, and in particular, using $K^\a_{ij} = 0$
for $i \not= j$  we get
$$ \wt{K}^\a_{ii} \ = \ \sum_{m} \ A_{m i}^T \, K^\a_{m m} \, A_{mi} \ \ \ (\mathrm{no \ sum \ on \ } i) \ . $$
The admitted $\La$ are thus identified as those built with the
$A_{ij}$ satisfying the conditions
\begin{eqnarray} \sum_{m} \ A_{m i}^T K^\a_{m m} A_{mj} & = & 0 \ \ \
\mathrm{for} \ i \not= j \ , \label{eq:lacond1} \\
\sum_{m} \ A_{m i}^T \, K^\a_{m m} \, A_{mi} & = & R_{\a \b} \, K^\b_{ii}
\ . \label{eq:lacond2} \end{eqnarray}

A discussion of solutions to
\eqref{eq:lacond1}, \eqref{eq:lacond2} is more conveniently
conducted in terms of infinitesimal generators; this requires
rather complex computations for a full analysis of the general
case, and these will be presented in a companion work
\cite{GReuclidean}.

\section{Characterization of canonical maps}
\label{sec:canmaps}

We will now discuss canonical transformations (see Definitions
\ref{sec:maps}.3 and \ref{sec:maps}.4 above); again
it is clear that these form a group, which
will be denoted as ${\tt Can} (M)$, or more precisely ${\tt Can}
(M, g,{\bf J})$. We will proceed as for hyperkahler maps, i.e.
first discuss the relation between the general case and the case
where the structure is in standard form at least at a given point,
and then discuss the characterization of canonical maps for
structures in standard form.

\subsection{Canonical maps for Euclidean versus general manifolds}

Let us again fix a reference point $x_0 \in M$ and perform the
change of coordinates $R$ which takes the Riemannian metric into
standard form at $x_0$, and subsequently the change of coordinates
$S$ which, leaving $g$ in standard form at $x_0$, takes the
complex structures $J_\a$ (and hence the symplectic forms
$\om_\a$) into standard form at $x_0$. Proceeding as in Section
\ref{sec:hkmaps}, we obtain the following.

\medskip\noindent
{\bf Lemma \ref{sec:canmaps}.1.} {\it Let $(M,g;{\bf J})$ be a
hyperkahler manifold; let $R(x)$ be the transformation taking $g$
and $J_\a$ into standard form $(g_0=\delta, \Jb_0)$ at the point $x \in M$.
The group of
canonical transformations {\rm at the point $x$} is given by $$ {\tt
Can} (g,\Jb) \ = \ R^{-1} (x) \ {\tt Can} (\de,\Jb_0) \ R(x)  \ ,
$$ where ${\tt Can} (\de,\Jb_0)$ is the group of canonical
transformations for $g$ and $\Jb$ in standard form.}

\subsection{Characterization for structures in standard form}

We have then to characterize the group ${\tt Can}_0 := {\tt Can}
(\de,\Jb_0)$ of maps which preserve $\om \w \om$ for standard \hK
structures; actually most of the discussion will be the same for
standard or generic form of these.

The key observation is that for any symplectic form $\om$ we can
write the volume form on any of the local four-dimensional
manifolds $U_a$ built in Section \ref{sec:maps} as
\beq\label{eq:partvol} \Om_{(a)} \ = \ \pm \, \iota_a^* [(1/2)
(\om \w \om)] \ , \eeq the sign depending on the orientation of
$\iota_a^* \om$. (In the same way, the volume form $\Om$ on the
$4n$-dimensional manifold $M$ can be written as $\Om = \pm
[(1/(2n!)) (\om \w ... \w \om)]$.)

In local coordinates we have $\om =  K_{ij} \d x^i \d x^j$ and hence
$$ \frac{1}{2} \ (\om \w \om) \ = \ \frac{1}{2} \
K_{ij} \, K_{\ell m} \ \d x^i \w \d x^j \w \d x^\ell \w \d x^m
\ . $$
Under a map with Jacobian $\La$, $K$ is transformed into
$\wt{K} = \La^T K \La$; correspondingly, the form $(1/2) (\om \w
\om)$ is rewritten as $$ \frac{1}{2} (\wt{\om} \w \wt{\om} )  \ = \
\frac{1}{2} \ ( \wt{K}_{ij} \wt{K}_{\ell m} ) \ \d x^i \w \d x^j \w \d
x^\ell \w \d x^m \ . $$

When we look at the volume form on $U_a$, only coordinates
$i,j,\ell,m$ in the range $\mathcal{R}_a := [4(a-1) + 1 ,...,4a]$ should appear; in other words, the operation $\iota_a^*$ sets to zero all four-forms $\d
x^i \w \d x^j \w \d x^\ell \w \d x^m$ except those with exactly
(any permutation of) the four suitable coordinates.

Thus we have, with $i,j,\ell,m \in \mathcal{R}_a$,
$$ V_{(a)} \ = \ \frac{1}{2} \, (\eps_{ij\ell m} K_{ij} K_{\ell m} ) \
\Om_{(a)} \ ; \ \ \wt{V}_{(a)} \ =  \frac{1}{2} \, (\eps_{ij\ell
m} \wt{K}_{ij} \wt{K}_{\ell m} ) \ \Om_{(a)} \ . $$ The central
object is thus the quantity \beq \P_2 (K) \ := \ (1/8) \
(\eps_{ij\ell m} K_{ij} K_{\ell m} ) \eeq (see Section
\ref{sec:bkI} and Appendix A); and a map is canonical if and only
if \beq\label{eq:cc} \sum_a \iota_a^* [ \P_2 (\tilde{K})] \ \equiv \
\sum_a \iota_a^* [\P_2 (\La^T K \La)] \ = \ \sum_a \iota_a^* [\P_2
(K)] \eeq for any $K$ corresponding to a symplectic form $\om \in
\S$.

It should be noted that -- as easy to check, e.g. by direct
computation (see also Appendix A) -- for a generic antisymmetric
matrix $K$ it results \begin{eqnarray} \label{eq:de2det} \P_2
(\La^T K \La) &=& \P_2 (K) \ \Det (\La ) \ .
\end{eqnarray}

It will again be convenient to write the matrix $\La$, as well as
the $K=K_\a$, in terms of four-dimensional blocks; we will use the
notation set up in Section \ref{sec:hkmaps}. With this, it turned
out that $\wt{K}_{ij} = A_{\ell i}^T K^0_{\ell m} A_{mj}$, and in
particular, using $K^0_{ij} = 0$ for $i \not= j$  (no sum on $i$)
$$ \wt{K}_{ii} \ = \ \sum_{m} \ A_{m i}^T \, K^0_{m m} \, A_{mi} \ . $$

Thus, the condition to have a canonical transformation
\eqref{eq:cc} reads now \beq\label{eq:cc2a} \sum_i \ \P_2 \[
\sum_m \, A_{m i}^T \, K^0_{m m} \, A_{mi} \] \ = \ \sum_i \ \P_2
\[ K^0_{mm} \] \ ; \eeq using now $\P_2 [ K^0_{mm} ] = 1$, this is
also written as \beq\label{eq:cc2} \frac{1}{n} \ \sum_i \ \P_2 \[
\sum_m \, A_{m i}^T \, K^0_{m m} \, A_{mi} \] \ = \ 1 \ . \eeq

This can be written in a slightly different form by introducing the
notation
\beq \P_1 (A,B) \ = \ (1/8) \ \epsilon_{ijkm} \, A_{ij} \, B_{km} \ ; \eeq
it is easily checked that $\P_1(B,A) = \P_1 (A,B)$ and $\P_2 (A) =
\P_1 (A,A)$. Moreover,
\begin{eqnarray*} \P_2 (A + B) &=& \P_1 (A,A) + \P_1 (A,B) + \P_1 (B,A) + \P_1 (B,B) \\
 &=& \P_2 (A) + \P_2 (B) \ + \ 2 \, \P_1 (A,B) \ . \end{eqnarray*}
Using this, condition \eqref{eq:cc2} can be rewritten as
\beq\label{eq:cc3}    \frac{1}{n} \sum_{i=1}^n \ \[ \sum_m \P_2 (
A_{mi}^T \, K^0_{mm} \, A_{mi} ) \ + \ 2 \ \sum_{m,\ell} \, \P_1
\( A_{mi}^T K^0_{mm} A_{mi} , A_{\ell i}^T K^0_{\ell \ell} A_{\ell
i} \) \] \ = \ 1 \ . \eeq We can summarize our discussion as
follows.

\medskip\noindent
{\bf Lemma \ref{sec:canmaps}.2.} {\it The group ${\tt Can}_0$ is
the group of all the matrices $\La$ of the form \eqref{eq:lambda}
which satisfy \eqref{eq:cc3}.}

\section{Dual \hK structures}
 \label{sec:dual}

In Euclidean space we have two kinds of standard \hK and \hS
structures, characterized by their different orientation, as
recalled in Section \ref{sec:stand_sub}. Both of these are needed
when we want to describe Dirac mechanics in hyperhamiltonian terms
(in this frame they are associated to opposite helicity states)
\cite{GR}.

\subsection{Construction of dual structures and standard forms}

We want to discuss briefly the relation between the map taking a
\hK structure into standard form and its action on the associated
\hK structure of opposite orientation. The construction of Section
\ref{sec:maps} allows to essentially reduce the discussion to the
four-dimensional case.

The orientation-reversing map on $\T M$ at the reference point
$x_0$ can be described in terms of a block diagonal matrix $R_0$
satisfying $R_0^T = R_0^{-1} = R_0$ and $\Det (R_0) = - 1$. In
coordinates, the simplest such map can be either the reversing of
a coordinate axis (say the first one), or the exchange of two
coordinate axes (say the first two); we will refer to these as
reversing and parity-reversing maps respectively.  In the first
case we write it as $R_0 = \rho_0$, while in the second one we
write $R_0=\eta_0$ (in each block); that is,
$$ \rho_0 \ = \ \pmatrix{-1&0&0&0\cr 0&1&0&0\cr 0&0&1&0\cr 0&0&0&1\cr}
 \ ; \ \ \eta_0 \ = \ \pmatrix{0&1&0&0\cr 1&0&0&0\cr 0&0&1&0\cr 0&0&0&1\cr} \ . $$
Needless to say, if we want to change orientation in several (or
all) of the four-dimensional blocks, the operation should be
applied to each of these.

The $R$ defined on $M$ is identified by $R (x_0) = R_0$ and by the
requirement that $\nabla R = 0$. The metric $g$ and hence the
Levi-Civita connection $\nabla$ and its coefficients $A_i$ are
invariant under $R$.

As $R$ is covariantly constant (and thus are $J_\a$ and $\om_\a$),
it follows immediately that the transformed complex structures
$\wt{J}_\a$ (respectively, symplectic structures $\wt{\om}_\a$)
are also covariantly constant.

Let us now consider a \hK structure $(g^0,\Jb^0)$ on $M$, and take
it into standard form $(g,\Jb)$ -- with positive orientation --
via a map $\Phi$ with Jacobian $\La$, as discussed in previous
sections.

\medskip\noindent
{\bf Lemma \ref{sec:dual}.2.} {\it The map $R_0$ takes $(g,\Jb)$
into a \hK structure $(\ol{g},\ol{\Jb})$ which is also in standard
form but with negative orientation.}

\medskip\noindent
{\bf Proof.} In order to check that $(\ol{g},\ol{\Jb})$ still
provides a \hK structure on $M$, it suffices to check that the
$\ol{J}_\a$ are covariantly constant under the connection $\ol{\nabla}$
corresponding to $\ol{g}$ and satisfy
the quaternionic relations. The first fact follows from the
previous observation, and the second from the tensorial nature of
the $J_\a$. Finally, it is obvious that $R$ changes orientation.
We should still check that the $\ol{J}_\a$ are in standard form
(with reversed orientation); this follows easily from an explicit
computation at the reference point $x_0$.
Note also that $\ol{g} = g$ (and hence $\ol{\nabla} = \nabla$),
as $R$ is orthogonal. \EOP
\bigskip

Inverting the map $\Phi$ with Jacobian $\La$ (i.e. considering the
map $\Phi^{-1}$ with Jacobian $\La^{-1}$) we now get
$(\wt{g}^0,\wt{\Jb}^0)$, given explicitly by \beq \wt{g}^0 \ = \
(\La^T)^{-1} \, \wt{g} \, (\La)^{-1} \ = \ g^0 \ \ ; \ \ \
\wt{J^0}_\a \ = \ \La \, \wt{J}_\a \, \La^{-1} \ . \eeq We say
that $(\wt{g}^0,\wt{\Jb}^0)$ is the \hK structure on $M$ {\bf
dual} to $(g^0,\Jb^0)$. Note that dual \hK structures share the
same Riemannian metric.

\subsection{Dirac structures}

On physical grounds -- e.g. in providing a hyperhamiltonian
description of the Dirac equation -- it is sometimes needed to
consider a pair of dual \hK structures.

\medskip\noindent
{\bf Definition \ref{sec:dual}.1.} {\it A pair of mutually dual
\hK structures ${\bf J}$ and ${\bf \^J}$ on $(M,g)$ is said to be
a {\rm Dirac structure} on $(M,g)$, and denoted as $({\bf J , \^J}
)$.}

\medskip\noindent
{\bf Remark \ref{sec:dual}.1.} A Dirac structure is
characterized not by a single unit sphere ${\bf S}$ in the
space ${\bf Q}$ of Kahler structures, but by a pair of dual
unit spheres; referring to their orientation we
will denote these by ${\bf S}_{+}$ and ${\bf S}_{-}$. \EOR

\medskip\noindent
{\bf Remark \ref{sec:dual}.2.} The discussion given here in terms
of \hK structures could have been performed in terms of \hS
structures; in this framework, we could speak of Dirac-symplectic
structures, and denote the unit spheres in $\mathcal{Q}$
characterizing such a structure as $\mathcal{S}_{+}$ and
$\mathcal{S}_{-}$. \EOR

\medskip\noindent
{\bf Remark \ref{sec:dual}.3.} In this sense, and in view of a
discussion of canonical maps, it is essential to note that -- as
apparent from the construction of dual \hK structures -- the splitting of
$\T M$ into four-dimensional invariant subspaces is the same for both
members of a pair of dual \hK structures.
In other words, these are also invariant subspaces
for the Dirac structure, and no distinction between the dual
structures can be made on the basis of the induced splitting.
\bigskip

\subsection{Hypersymplectic transformations for Dirac structures}

Consideration of Dirac structures calls for a discussion of their
canonical and (the equivalent of their) hyperkahler
transformations. While in the former case the definition can be
extended unaltered from the \hK case, in the latter we will need a
slight generalization in order to consider both structures at the
same time and allow some mixing.

\medskip\noindent
{\bf Definition \ref{sec:dual}.2.} {\it Let $(M,g)$ be a real
Riemannian manifold of dimension $4n$, and let $(\Jb^{(1)} ,
\^\Jb^{(1)})$, $(\Jb^{(2)} , \^\Jb^{(2)})$, be two Dirac
structures in it. We say that these are equivalent if $\Jb^{(1)}$
is equivalent to $\Jb^{(2)}$ and $\^\Jb^{(1)}$ is equivalent to
$\^\Jb^{(2)}$.}

\medskip\noindent
{\bf Definition \ref{sec:dual}.3.} {\it Let $(M,g)$ be a real
Riemannian manifold of dimension $4n$, equipped with a Dirac
structure $(\Jb , \^\Jb)$. We say that the orthogonal map $\phi :
M \to M$ is {\rm Dirac-hyperkahler} (or {\rm Dirac-quaternionic})
if it maps the Dirac structure into an equivalent one.
Equivalently, if its pullback $\phi^*$ satisfies $\phi^* :
{\mathcal S}_\pm \to {\mathcal S}_\pm$.}

\medskip\noindent
{\bf Remark \ref{sec:dual}.4.} We can also define {\it strongly
Dirac-symplectic} maps as those leaving the Dirac structure
invariant; that is, those for which $\phi^* (\om_a) = \om_a$,
$\phi^* (\^\om_a) = \^\om_a$ for $a=1,2,3$. The requirement for a
map to be strongly Dirac-symplectic is very restrictive, and in
general one should expect these maps, apart from trivial ones, to
be quite exceptional. \EOR

\medskip\noindent
{\bf Remark \ref{sec:dual}.5.} As for Dirac-hyperkahler maps, a
large class of them is provided by standard Hamiltonian flows
under any of the involved symplectic structures. It should be
stressed that if we consider the flow related to say $\om_1$, this
will be strongly hypersymplectic for the \hK structure with
reverse orientation (in that the $\^\om_a$ are left invariant, as
follows from $[Y_i,\^Y_j]=0$), and hypersymplectic for the \hK
structure to which $\om_1$ belongs. \EOR

\subsection{Canonical transformations for Dirac structures}

Let us now look at canonical transformations. As noted in Remark
\ref{sec:dual}.3 above, the invariant subspaces of $\T M$ are just
the same for two dual hypersymplectic structures, and are hence
attached to the full Dirac structure. Moreover, we have $\om_\a \w
\om_\a = \^\om_\b \w \^\om_\b$ (no sum on $\a$ and $\b$). In other
words, canonical transformations will be the same for dual \hK (or
hypersymplectic) structures, and these will also be the canonical
transformations for the corresponding Dirac structure.

We can then just rephrase our definition of canonical
transformation in the present framework, and give a formalization
of the above remark.

\medskip\noindent
{\bf Definition \ref{sec:dual}.4.} {\it Let $(M,g)$ be a real
Riemannian manifold of dimension $4n$, equipped with a Dirac structure
$\mathcal{D} = (\Jb , \^\Jb)$; let $\mathcal{S}_\pm$ be the corresponding
symplectic Kahler spheres. The map $\phi : M \to M$
is said to be {\rm canonical} for $\mathcal{D}$ if, for any
$\om \in \mathcal{S}_+$ and and $\^\om \in \mathcal{S}_-$, it preserves
the four-forms $\om \w \om$ and $\^\om \w \^\om$.}

\medskip\noindent
{\bf Lemma \ref{sec:dual}.3.} {\it Let Let $(M,g)$ be a real
Riemannian manifold of dimension $4n$, equipped with a Dirac
structure $\mathcal{D} = (\Jb , \^\Jb)$. The set $\mathtt{Can}
(g,\Jb , \^\Jb )$ of canonical transformations for it coincides
with the sets of canonical transformations for each of the
associated \hK structures, \beq \mathtt{Can} (g,\Jb , \^\Jb ) \ =
\ \mathtt{Can} (g,\Jb) \ = \ \mathtt{Can} (g, \^\Jb ) \ . \eeq}

\medskip\noindent
{\bf Proof.} The forms $\Om_+ = (1/2) (\om \w \om)$ and $\Om_- =
(1/2) (\^\om \w \^\om)$ (where $\om \in \mathcal{S}_+$, $\^\om \in
\mathcal{S}_-$) built from dual \hK structures are equal up to a
sign, $\Om_- = - \Om_+$; thus preservation of one of them implies
preservation of the other one as well. \EOR

\section{Canonical property of the Hyperhamiltonian flow}
\label{sec:hhcan}

We want now to show that the hyperhamiltonian vector fields, first
introduced in \cite{GM} (see also \cite{GM2,MT}) provide an
unfolding of canonical transformations, pretty much in the same
way as Hamiltonian vector fields in the case of symplectic
structures. We will first recall the basic definitions of
hyperhamiltonian vector fields, and then show they enjoy the
canonicity property.

\subsection{Hyperhamiltonian vector fields}
\label{sec:hh}

Let $(M,g;\Jb)$ be a hyperkahler manifold. Given a triple of
functions on $M$, $\overrightarrow{\h} = \{ \h^1,\h^2,\h^3)$,
these identify three hamiltonian vector fields via the standard
hamiltonian relation (no sum on $\a$) \beq\label{eq:hamilton} X_\a
\interno \om_\a \ = \ \d \, \h^\a \ . \eeq

The hyperhamiltonian vector field $X$ associated to the triple
$(\h^1 , \h^2 , \h^3 )$ is defined as the sum of the $X_\a$'s,
i.e. \beq\label{eq:hyperham} X \ := \ \sum_\a \, X_\a \ . \eeq

This was introduced -- and several of its properties discussed --
in \cite{GM}; see also \cite{GMspt} for a discussion of the
integrable case, \cite{GR} for some physical applications (to
systems with spin), and \cite{MT} for a complex analysis approach.

\medskip\noindent
{\bf Remark \ref{sec:hhcan}.1.} Equivalent hypersymplectic
structures will not generate the same hyperhamiltonian dynamics
for a given triple of Hamiltonians; needless to say, if we operate
corresponding rotations in the space ${\bf Q}$ {\it and } in the
space of Hamiltonians as well -- i.e. consider complex structures
$\wt{J}_\a = R_{\a \b} J_\b$ and hence symplectic forms
$\wt{\om}_\a = R_{\a \b} \om_\b$, and Hamiltonians $ \wt{\h}_\a =
R_{\a \b} \h_\b$ with the same $R \in SO(3)$ -- then we obtain the
same dynamics. \EOR
\bigskip

The results we want to prove are of local nature, i.e. we can work
on a single chart of the hyperkahler manifold $(M,g;\Jb)$. In the
following we will use local coordinates $x^i$, $i = 1, ... , 4n$.
With these (and recalling the notation introduced in Section
\ref{sec:bkI}), the hyperhamiltonian vector field $X$ will be
written as \beq X \ = \ f^i \ \pa_i \ ; \ \ \ \mathrm{where} \ \
f^i \ = \ \sum_\a \ (M_\a)^{ij} \, \pa_j \h^\a \ . \eeq

\subsection{Hyperhamiltonian flows and canonical transformations}
\label{sec:canflow}

Let us look at the transformations undergone by an arbitrary
symplectic form $\om \in \mathcal{S}$, and by the associated
volume form $V_a (\om) := (1/2) \iota_a^* (\om \w \om)$ on $U_a$,
under a hyperhamiltonian flow.

We will work in local coordinates around the reference point $x_0$
at which the metric and the \hS structure are in standard form. We
freely move the indices $\a,\b,...$ (referring to the \hK triple)
up and down for typographical convenience.

\subsubsection{Lie derivative of the symplectic forms.}

In order to know how the $\om_\a$ change under the
hyperhamiltonian flow we have to compute the Lie derivative $ \L_X
(\om)$.

In our case $\om$ is by definition closed (being a symplectic form),
hence we have $ \L_X \om = \d (X \interno \om)$.

We will use the shorthand notations (note $D_\a = D_\a^T$)
\beq\label{eq:PD}  P^\b_{\ k} \ := \ (\pa H^\b / \pa x^k) \ ; \ \ \
D^\a_{ij} \ := \ \frac{\pa^2 H^\a}{\pa x^i \pa
x^j} \ . \eeq

\medskip\noindent
{\bf Lemma \ref{sec:hhcan}.1.} {\it For $X$ the hyperhamiltonian
vector field, it results \beq\label{eq:lambda4} \L_X (\om_\a ) \ =
\ \epsilon_{\a \b \ga} \ [\pa_i (P^\b_{\ k} (Y_\ga)^k_{\ j} ) ] \d
x^i \w \d x^j \ . \eeq}

\medskip\noindent
{\bf Proof.}
We have $$ X \interno \om_\a  = (1/2) \, \(
K^\a_{ij} f^i \d x^j  - K^\a_{ij}  f^j  \d x^i \) \ . $$ By a
rearrangement of indices, and using $K_\a^T = - K_\a$, this is
also rewritten as \beq\label{eq:xintom} X \interno \om_\a \ = \
f^i \, K^\a_{ij} \, \d x^j \ := \ \la^\a_{\ j} \, d x^j \ ; \eeq
here we defined the quantities $\la^\a_{\ j} = f^i (K^\a)_{ij}$ on
the r.h.s. as they will appear repeatedly in the following. With
this, we get \beq\label{eq:der1} \L_X (\om_\a) \ = \ \d \la^\a_j
\w \d x^j \ = \ (\pa_i \la^\a_j) \ \d x^i \w \d x^j \ . \eeq

These hold for a generic vector field; now we specialize to the
hyperhamiltonian case. With the notation \eqref{eq:PD},
the hyperhamiltonian vector field is
given by \beq f^m \ = \ M_\b^{mk} \, \pa_k H^\b \ = \ P^\b_{\ k}
(M_\b^T)^{km} \ = \ - \, P^\b_{\ k} M_\b^{km} \ . \eeq
 A simple computation shows that
\begin{eqnarray}
\lambda^\a_j &=& - P^\b_{\ k} M_\b^{km} K^\a_{mj} \ = \
 - \, P^\b_{\ k} \, (Y_\b)^k_{\ m} \, (Y_\a)^m_{\ j} \nonumber \\
&=& - \ P^\b_{\ k} \ \[ \epsilon_{\b \a \ga} (Y_\ga)^k_{\ j}  \ - \ \de_{\b \a} \de^k_{\ j} \] \ = \
 P^\a_{\ j} \ + \ \epsilon_{\a \b \ga} \, P^\b_{\ k}
(Y_\ga)^k_{\ j} \ . \label{eq:lambda2} \end{eqnarray}

Then \eqref{eq:lambda4} follows at once recalling that $\L_X (\om_\a) \
= \ \d \la^\a_j \w \d x^j $ and $\pa_i P^\a_{\ j} = D^\a_{ij}$.
Indeed, differentiating \eqref{eq:lambda2} we get
\beq\label{eq:lambda3} d \la^\a_j \ = \ D^\a_{i j} \, \d x^i \ + \
\epsilon_{\a \b \ga} \, [\pa_i (P^\b_{\ k} (Y_\ga)^k_{\ j})] \d
x^i \ ; \eeq the first term does not contribute to the final
result since $D^\a_{ij} \d x^i \w \d x^j = 0$ due to $D_\a^T =
D_\a$. \EOP

\medskip\noindent
{\bf Remark \ref{sec:hhcan}.2.} For a generic form $\om = c_\a
\om_\a \in \mathcal{S}$ (hence with $|c| = \sum c^2_\a = 1$) we
obtain easily that
$$ \L_X (\om \w \om) \ = \ c_\eta \cdot c_\eta \, \epsilon_{\a \b \ga} \
\[ \pa_i \( P^\b_{\ q} (Y_\ga)^q_{\ j} \) \] \, K^\a_{\ell m} \
\d x^i \w \d x^j \w \d x^\ell \w \d x^m \ . $$
If $\pa_i Y^\ga = 0$ (which is verified in the Euclidean case) we
get simply (no sum on $\a$)
$$ \L_X (\om_\a  \w \om_\a ) \ = \
\epsilon_{\a \b \ga} \,  K^\a_{\ell m} D^\b_{iq} (Y_\ga)^q_{\ j}   \
\d x^i \w \d x^j \w \d x^\ell \w \d x^m \ . \eqno{\odot} $$

\subsubsection{Lie derivative of the volume forms $\Om_a$.}
 Now we consider volume forms $\Om_a$ in the $U_a$. The key
observation here is that $ \Om_a = \d x^{4 a -3} \w \d x^{4 a - 2}
\w \d x^{4 a -1} \w \d x^{4 a}$ (no sum on $a$), can be written as
\beq\label{eq:volUa} \Om_a \ = \ \s_a (\om) \ \iota^*_a \[ (1/2) \
(\om \w \om ) \] \eeq for any unimodular $\om$, where $\s_a (\om)
= 1$ for $\iota^*_a \om \in \mathcal{S}$ and $\s_a (\om) = -1$ for
$\iota^*_a \om \in \widehat{\mathcal{S}}$ (that is, $\s_a (\om) =
\P [\iota_a^*(\om )] $).

Any symplectic form in four dimensions is written at the reference
point $x_0$ as the sum of the standard positively and negatively
oriented ones (this just follows from $Y_i$ and $\^Y_i$ being a
basis for the set of all the possible antisymmetric matrices in
dimension four); thus we may set \beq\label{eq:iota1} \iota^*_a
(\om ) \ = \ c_\a \, \om_\a \ + \ \^c_\a \, \^\om_\a \ . \eeq
It follows from a standard explicit computation that for such  $\om$,
$$ \iota^*_a (\om \w \om) \ = \ \iota_a^* \[ \sum_\a \ \[ c_\a^{\, 2} \
(\om_\a \w \om_\a ) \ + \ \^c_\a^{\, 2} \ (\^\om_\a \w \^\om_\a )
\] \] \ , $$ with exactly the same $c_\a$ and $\^c_\a$ as above;
in fact, it is easy to check that $\om_\a \w \om_\b = 0$ for $\a
\not= \b$, and that $\om_\a \w \^\om_\b = 0$ for all $\a$ and
$\b$. (Needless to say, for $\om \in \mathcal{S}$
only the $c_\a$ are nonzero, and conversely for $\^\om \in
\^\S$.)

We also recall that for symplectic forms (or complex structures) in
standard form, all the matrix elements $K_{ij}$ (or $Y^i_{\ j}$)
with $i$ and $j$ not belonging to the same four-dimensional block
are zero.

Equation \eqref{eq:lambda4} yields (no sum on $\a$)
\beq \label{eq:lxomom}    \L_X (\om_\a  \w \om_\a ) =
\epsilon_{\a \b \ga}  \, K^\a_{\ell m} \( D^\b_{iq} (Y_\ga)^q_{\ j}  +  P^\b_{\ q} \pa_i (Y_\ga)^q_{\ j} \) \, \d x^i \w \d x^j \w \d x^\ell \w \d x^m \ . \eeq
For a general $\om = c_\eta \om_\eta$, this provides
\begin{eqnarray} \label{eq:lxomomgen}    \L_X (\om  \w \om ) &=&
(c_\eta \cdot c_\eta) \epsilon_{\a \b \ga}  \, K^\a_{\ell m} \( D^\b_{iq} (Y_\ga)^q_{\ j} +  P^\b_{\ q} \pa_i (Y_\ga)^q_{\ j} \) \,  \times \\
& & \ \times \ \d x^i \w \d x^j \w \d x^\ell \w \d x^m \ . \nonumber
\end{eqnarray}
Here everything can be computed by evaluating matrices at the
single reference point $x_0$, except the derivative $\pa_i
(Y_\ga)^q_{\ j}$. However, this can also be transformed into an
algebraic quantity by recalling that $\nabla J_\ga = 0$. In
coordinates, this reads \beq\label{eq:eqA} \pa_i (Y_\ga) \ + \
[A_i , Y_\ga] \ = \ 0 \ ; \eeq here $A_i$ is the connection
matrix, defined by $(A_i)^j_{\ k} = \Gamma^j_{ik}$, with
$\Gamma^j_{ik} = \Gamma^j_{ki}$ the Christoffel symbols for the
metric $g$.

Using \eqref{eq:eqA} allows to rewrite \eqref{eq:lxomomgen} as
\begin{eqnarray*}    \L_X (\om \w \om) &=& (c_\eta \cdot c_\eta )
\epsilon_{\a \b \ga}
\[ K^\a_{\ell m} \( D^\b_{iq} (Y_\ga)^q_{\ j} \ + \
P^\b_{\ q} ( (A_i)^q_{\ m} (Y_\ga)^m_{\ j} - (Y_\ga)^q_{\ m}
(A_i)^m_{\ j} ) \) \] \ \times \label{eq:lxomom2} \\ & & \ \times
\, \d x^i \w \d x^j \w \d x^\ell \w \d x^m \ .
\end{eqnarray*}

\medskip\noindent
{\bf Remark \ref{sec:hhcan}.3.} In all these formulas, the action
of $\iota_a^*$ amounts to setting to zero all variables (and its
differential) not belonging to the $a$-th block. \EOR
\bigskip

We are now ready to complete our computations; we will set their
results (respectively, for the case of constant $Y$ and the
general case) in the form of a Lemma (for the special case of
constant $Y$) and a Theorem for the general case. (We also
discuss, in the Appendix B, an alternative -- combinatorial --
approach to the proof of our main result in Theorem
\ref{sec:hhcan}.1.)

\medskip\noindent
{\bf Lemma \ref{sec:hhcan}.2.} {\it In the case where, for all
$\ga$ and all $i$, $\pa_i Y_\ga = 0$, any hyperhamiltonian flow
preserves $\iota_a^* (\om \w \om)$ and hence the volume forms
$\Om_a$ on $U_a$.}

\medskip\noindent
{\bf Proof.} In the case $\pa_i Y^\ga = 0$, formula
\eqref{eq:lxomomgen} reduces to \beq\label{eq:lxomomgeneucl} \L_X
(\om  \w \om ) \ = \ (c_\eta \cdot c_\eta) \, \epsilon_{\a \b \ga} \,
K^\a_{\ell m} D^\b_{iq} (Y_\ga)^q_{\ j}  \ \d x^i \w \d x^j \w
\d x^\ell \w \d x^m \ . \eeq It suffices to write down this, or
more precisely its pullback under $\iota_a^*$, in explicit terms;
for ease of notation we will consider the case $a=1$, so that only
variables $\{x^1,...,x^4\}$ are nonzero (any function should be
considered as evaluated with $x^k = 0$ for $k > 4$). We get
$$
\begin{array}{l}
(1/2) \, \L_X \( \om \w \om \) \ = \\
\ = \ \{ c_1^2 \ \[ \( D^2_{14} + D^2_{23}- D^2_{32} - D^2_{41} \) \, + \, \( D^3_{13}  - D^3_{24} - D^3_{31} + D^3_{42} \) \]  \\
\ + \ c_2^2 \ \[ \( D^1_{12} - D^1_{21} + D^1_{34} - D^1_{43} \) \, + \, \( D^3_{13} - D^3_{24} - D^3_{31} + D^3_{42} \) \]  \\
\ + \ c_3^2 \ \[ \( D^1_{12} - D^1_{21} + D^1_{34} - D^1_{43} \)
\, + \, \( D^2_{14} + D^2_{23} - D^2_{32} - D^2_{41} \) \] \} \
\times \\
\ \times \d x^1 \w \d x^2 \w \d x^3 \w \d x^4 \ .
\end{array} $$
Recalling that $D^\a_{ij} = D^\a_{ji}$ we conclude
that each of the coefficients of the $c_\a^2$ vanish separately,
hence $\L_X (\om \w \om ) = 0$ as stated. \EOP

\medskip\noindent
{\bf Theorem \ref{sec:hhcan}.1.} {\it Any hyperhamiltonian flow
preserves $\iota_a^* (\om \w \om)$ and hence the volume forms $\Om_a$
on the $U_a$.}

\medskip\noindent
{\bf Proof.} The variation of $\iota_a^* (\om \w \om) $ under $X$
is given by $\iota_a^* [ \L_X (\om \w \om)]$. To evaluate this we
make use of \eqref{eq:lxomom2} and of Remark \ref{sec:canflow}.3;
moreover, for ease of notation, we will focus on $a=1$.
That is, we should compute \eqref{eq:lxomom2} with all $i,k,\ell,m$
indices restricted to the range $1,...,4$.

We note that according to \eqref{eq:lxomom2}, $\L_X (\om \w \om)$,
and therefore $\iota_a^* [ \L_X (\om \w \om)]$ as well, is
the sum of two terms; these correspond respectively to $ K^\a_{\ell
m} D^\b_{iq} (Y_\ga)^q_{\ j}$ and to $K^\a_{\ell m} P^\b_{\ q}
[A_i,Y_\ga]^q_{\ j} $. The first term is exactly the one which
was already evaluated in the case $\pa_i Y_\ga=0$; it vanishes
as stated by Lemma \ref{sec:hhcan}.2 (and shown in its proof).

We therefore have only to show that \begin{eqnarray}    & & (c_\a
/2) \epsilon_{\a \b \ga} \iota_a^* \[ K^\a_{\ell m} P^\b_{\ q}
\( (A_i)^q_{\ p} (Y_\ga)^p_{\ j} - (Y_\ga)^q_{\ p} (A_i)^p_{\ j}
\) \ \times \right. \nonumber \\
 & & \left. \d x^i \w
\d x^j \w \d x^\ell \w \d x^m \]   \label{eq:thm} \\
& := & (c_\a /2) \  \epsilon_{\a \b \ga} \ \Theta^{(a)}_{\a \b
\ga}  \nonumber  \end{eqnarray} vanishes; here we have of course
defined
$$ \Theta^{(a)}_{\a \b \ga } \ = \ \iota_a^* \[ K^\a_{\ell m} P^\b_{\ q} \( (A_i)^q_{\ p} (Y_\ga)^p_{\ j} - (Y_\ga)^q_{\ p}  (A_i)^p_{\ j} \)  \d x^i \w
\d x^j \w \d x^\ell \w \d x^m \] \ . $$ Note that here the
coefficients $c_\a$ (satisfying $|c|^2 = 1$) and the vectors
$P^\b$ are completely arbitrary; thus the r.h.s. of \eqref{eq:thm}
should vanish for any choice of these. In other words we should
have $\Theta^{(a)}_{\a \b \ga} = 0$ for all choices of the indices
$\a , \b , \ga$, provided these are all different, $\a \not= \b
\not= \ga \not= \a$.

The expression for $\Theta$ only  involves quantities computed at
the reference point $x_0$, and we can hence make use of the
explicit expressions for the standard form of the $Y_\a$ and the
$K_\a$.

Using these, choosing $a=1$ (and omitting the index $a$), and the
case of positive orientation in the first block, we get e.g.
\begin{eqnarray*}
\Theta_{123} &=& - 2 \, \{ [(\Ga^1_{14} + \Ga^1_{23} - \Ga^1_{32} - \Ga^1_{41}) - ( \Ga^3_{12} - \Ga^3_{21} + \Ga^3_{34} - \Ga^3_{43})] \ P^2_{\ 1} \\
& & \ + [(\Ga^2_{14} + \Ga^2_{23} - \Ga^2_{32} - \Ga^2_{41}) + ( \Ga^4_{12} -
  \Ga^4_{21} + \Ga^4_{34} - \Ga^4_{43}) ] \ P^2_{\ 2} \\
   & & \ + [(\Ga^1_{12} - \Ga^1_{21} + \Ga^1_{34} - \Ga^1_{43}) + (\Ga^3_{14} +
  \Ga^3_{23} - \Ga^3_{32} - \Ga^3_{41})] \ P^2_{\ 3} \\
   & & \ + [ (\Ga^2_{21} -\Ga^2_{12} + \Ga^2_{43} - \Ga^2_{34}) + (\Ga^4_{14} +
  \Ga^4_{23} - \Ga^4_{32} - \Ga^4_{41})] \ P^2_{\ 4} \} \ \times \\
   & & \ \times \ \d x^1 \w \d x^2 \w \d x^3 \w \d x^4 \ ; \end{eqnarray*}
recalling the property $\Ga^i_{jk} = \Ga^i_{kj}$, valid for any
Riemannian metric $g$, it is immediately seen that $\Theta_{123}$
vanishes. The same holds for all forms $\Theta_{\a \b \ga}$ with
$\a \not= \b \not= \ga \not= \a$ (explicit formulas are omitted
for the sake of brevity). This concludes the proof for positive
orientation. As usual, computations are the same up to certain
signs for negative orientation as well, and lead to the same
result. \EOP

\medskip\noindent
{\bf Remark \ref{sec:hhcan}.4.} The result of the Theorem above
can be restated as follows: {\it Any hyperhamiltonian flow
corresponds to a one-parameter family of canonical transformations
for the underlying hyperkahler structure}. \EOR

\medskip\noindent
{\bf Remark \ref{sec:hhcan}.5.} We have shown that the
hyperhamiltonian flow is canonical for the underlying \hK
structure (i.e. the one defining it through \eqref{eq:hyperham});
it turns out it is also canonical for the dual one. Indeed, the
structures $\om_\a$ and $\^\om_\a$ define the same invariant
subspaces in $\T_x M$ and define on these volume forms which only differ by a
sign, $\Om_a = - \^\Om_a$; it is thus a triviality that
preservation of the volume forms $\Om_a$ for the defining
structure entails preservation of the $\Om_a$ for the dual one. In
other words, $\L_X (\om \w \om) = 0$ implies $\L_X (\^\om \w
\^\om) = 0$; this follows at once from $(\^\om \w \^\om ) = - (\om
\w \om)$.  \EOR

\section{Examples: four dimensional Euclidean space}
\label{sec:examples}

We will discuss in detail hyperkahler and canonical maps for flat
\hK structures in a companion paper \cite{GReuclidean}; in this
section we will just discuss the simplest case of Euclidean space,
$M = \R^4$ with Euclidean metric $g(x) = \de$.

\subsection{Hyperkahler maps}

We have $(M,g)=(\R^4,\de)$ and either one of the standard \hK
structures (see Section \ref{sec:bkI}), to which we can always
reduce; we will for short just focus on the $Y_\a$, the situation
being completely analogous for the $\^Y_\a$.

To preserve the metric we are bound to consider orthogonal
transformations, i.e. $O(4)$. Moreover, we have to preserve
orientation, which ensures $\hSp (4) \sse SO(4)$.

The six generators of the Lie algebra $so(4) \simeq su(2) \oplus
su(2)$ can be chosen to be exactly $\{ Y_\a ; \^Y_a \}$. It is
immediate to check that the $\^Y_a$ (each of them commutes with
all of the $Y_a$) generate strongly hyperkahler transformations,
while the $Y_a$ themselves generate (non-strongly) hyperkahler
ones.

In a somewhat more detailed way, let us write a generic complex
structure $J \in {\bf S}$ as $J = \sum_a k_a Y_a$, where $k_\a$
are real constants and $|k|^2 := \sum_a k_\a^2 = 1$. A generic
element $\la$ of the algebra $so(4)$ will be written as $\la =
p_\a Y_\a + q_\a \^Y_\a$, where $p_\a,q_\a \in \R$. The
infinitesimal action of $\la$ on $J$ is given by
\begin{eqnarray*} J \ \to \ J' &=& J \ + \ \varepsilon \ [\la,J] \ = \ J
\ + \ \varepsilon \left( p_\a [Y_\a , J] \ + \ q_\a [\^Y_\a , J]
\right) \\ &=& J \ + \ \varepsilon \left( p_\a [Y_\a , J] \right)
\ = \ k_\b \, Y_\b \ + \ \varepsilon \ p_\a k_\b \, [Y_\a,Y_\b] \\
 &=& k_\b \, Y_\b \ + \ 2 \varepsilon \ \epsilon_{\a \b \ga} p_\a k_\b Y_\ga \ = \
(k_\ga + 2 \varepsilon \epsilon_{\a \b \ga} p_\a k_\b ) \ Y_\ga \\ &:=& z_\ga \
Y_\ga . \end{eqnarray*}

It is obvious that $J'$ is in the linear span of $(Y_1,Y_2,Y_3)$;
to check we are indeed on the unit sphere, it suffices to recall
we have to consider orthogonal transformations. We can also
compute explicitly (at first order in $\varepsilon$)
$$ |z|^2 \ = \ z_\a \, z_\a \ = \ k_\a k_\a +
2 \varepsilon k_\a \, \epsilon_{\b \ga \a} p_\b k_\ga +  O(\varepsilon^2 )
\ = \ k_\a k_\a \ + \ O(\varepsilon^2 ) \ . $$

In conclusion, as stated above, $\hSp (4) \simeq SO(4)$; more
precisely, all maps in the group $SO(4) \simeq SU(2) \times SU(2)$
generated by the $\{ Y_\a , \^Y_\a \}$ are hyperkahler, and those in the $SU(2)$ factor generated by the $\^Y_\a$ are strongly
hyperkahler.

In arbitrary $4n$ dimension, the invariance group will be still the direct product of two groups corresponding to hyperkahler and strong hyperkahler transformations. In accordance with general results on manifolds with special holonomy \cite{Be55} the invariance group will be $Sp(1)\times Sp(n)$, which reduces for $n=1$ ($4$-dimensional case) to $Sp(1)\times Sp(1)$ which is isomorphic to the group we have obtained here.

\subsection{Hyperhamiltonian flows and canonical transformations}

Let us now consider $(\R^4,\de)$ with standard \hK structure (with
positive orientation) from the point of view of canonical maps.
The volume form is just
$$ \Omega \ = \ \d  x^1  \w   \d  x^2  \w  \d  x^3  \w \d  x^4 \ . $$
We consider an arbitrary $\om \in \S$, i.e. $\om = c_\alpha
\om_\alpha$ with $|c|^2 = c_1^2 + c_2^2 + c_3^2 = 1$; for this we
have $(1/2) (\om  \w \om) = \Omega$. For a vector field $X = f^i
\pa_i$ it follows from standard computations (using also $|c|^2 =
1$) that
\begin{eqnarray} (X \interno \om) \,  \w   \, \om &=&
 f^1 \, \d  x^2  \w   \d  x^3  \w   \d  x^4 \ - \
f^2 \d  x^1  \w   \d  x^3  \w   \d  x^4 \nonumber \\
& & \ + \ f^3 \d  x^1  \w   \d  x^2  \w \d  x^4 \ - \ f^4
 \d  x^1  \w   \d  x^2  \w \d  x^3 \ .
\end{eqnarray} Specifying now that $X$ is the hyperhamiltonian
vector field corresponding to hamiltonians $\{ \h_1 , \h_2 , \h_3
\}$, see Section \ref{sec:hh}, we get
\begin{eqnarray}
(X\interno \om)  \w   \om &=&
(\pa_2 \h_1 + \pa_4 \h_2 + \pa_3 \h_3) \,
\d  x^2  \w   \d  x^3  \w   \d  x^4 \nonumber \\
 & & \ + \ (\pa_1 \h_1 - \pa_3 \h_2 + \pa_4 \h_3) \,
 \d  x^1  \w   \d  x^3  \w   \d  x^4 \nonumber \\
& & \ + \ (\pa_4 \h_1 - \pa_2 \h_2 - \pa_1
\h_3) \, \d  x^1  \w   \d x^2  \w   \d  x^4 \nonumber \\
& & \ + \ (\pa_3 \h_1 + \pa_1 \h_2 - \pa_2 \h_3) \,
 \d  x^1  \w   \d  x^2  \w   \d  x^3 \ .
\end{eqnarray} It follows from this that
\begin{equation} {\mathcal L}_X (\om  \w   \om ) \ = \
\d [ (X \interno \om)  \w   \om] \ = \ 0 \ . \end{equation}

In other words, we have shown by explicit computation that $\Omega
= (1/2) (\om  \w   \om) $ is preserved under {\it any }
hyperhamiltonian flow. (Actually our computation showed this only
for positively oriented \hS structures; the computation goes the
same way for negatively oriented ones.)
\bigskip

Let us go back to considering ${\mathcal L}_X (\om)$; using the
explicit expression for the hyperhamiltonian vector field, it
turns out by a direct computation that this can be written as
\begin{equation} {\mathcal L}_X (\om) \ = \ \frac12 \
\left( p_\alpha \, \om_\alpha \ + \ q_\alpha \, \^\om_\alpha \right) \end{equation}
with coefficients $p_\alpha , q_\alpha$ given by (here $\triangle$
is the Laplacian)
\begin{eqnarray*}
p_1 &=&  c_2 \triangle \h_3 - c_3 \triangle \h_2  \ , \
p_2 =  c_3 \triangle \h_1 - c_1 \triangle \h_3  \ , \
p_3 =  c_1 \triangle \h_2 - c_2 \triangle \h_1  \ ; \\
q_1 &=&  c_1 \left[ \(\pa_1^2\h_2 - \pa_2^2\h_2 +
\pa_3^2\h_2 -\pa_4^2 \h_2 \) \ - \ 2 \, \(\pa_1 \pa_2\h_3 +\pa_3 \pa_4 \h_3 \) \right] \\
& & \ - \ c_2 \left[ \( \pa_1^2 \h_1 - \pa_2^2\h_1 +
\pa_3^2\h_1 -\pa_4^2 \h_1\) \ - \ 2 \, \(\pa_1 \pa_4\h_3 - \pa_2 \pa_3 \h_3 \) \right] \\
& & \ + \ 2 \, c_3 \left[ \( \pa_1 \pa_2 \h_1 +\pa_3 \pa_4 \h_1 \) \ + \ \( \pa_1 \pa_4\h_2 - \pa_2 \pa_3\h_2 \) \right] \ , \\
q_2 &=& c_1 \left[ \( \pa_1^2\h_3 - \pa_2^2 \h_3-
\pa_3^2 \h_3 +\pa_4^2 \h_3 \) \ + \ 2 \, \( \pa_1 \pa_2 \h_2 - \pa_3 \pa_4\h_2 \) \right] \\
& & \ - \ c_3 \left[ \( \pa_1^2 \h_1 -\pa_2^2 \h_1-
\pa_3^2  \h_1+\pa_4^2 \h_1 \) \ - \ 2 \, \( \pa_1 \pa_3  \h_2+ \pa_2 \pa_4\h_2 \) \right] \\
& & \ - \ 2 \, c_2 \left[ \( \pa_1 \pa_2 \h_1 -\pa_3 \pa_4  \h_1\) \ + \ \( \pa_1 \pa_3  \h_3+ \pa_2 \pa_4  \h_3\) \right] \ , \\
q_3 &=& c_2 \left[ \( - \pa_1^2 \h_3 - \pa_2^2 \h_3 +
\pa_3^2 \h_3 + \pa_4^2\h_3 \) \ + \ 2 \, \( \pa_1 \pa_4\h_1 + \pa_2 \pa_3 \h_1 \) \right] \\
& & \ + \ c_3 \left[ \(  \pa_1^2\h_2 + \pa_2^2 \h_2 -
\pa_3^2 \h_2 - \pa_4^2\h_2 \) \ + \ 2 \, \( \pa_1 \pa_3 \h_1 - \pa_2 \pa_4  \h_1\) \right] \\
& & \ - \ 2 \, c_1 \left[ \( \pa_1 \pa_4 \h_2 + \pa_2 \pa_3 \h_2
\) \ - \ \( \pa_1 \pa_3 \h_3 -\pa_2 \pa_4\h_3 \) \right] \ .
\end{eqnarray*}

The essential point here is that -- as these explicit formulas
show -- the Lie derivative ${\mathcal L}_X (\om)$ of a symplectic
form $\om \in \mathcal{S} \subset \mathcal{Q}$ has components
along $\hat{\mathcal{Q}}$, i.e. the negatively-oriented forms.

This shows that in general the hyperhamiltonian flow, even in this
simple case, is canonical but {\it not} hyperkahler; see also
Remark \ref{sec:canmaps}.2.

An exception is provided by the choice
$ \h_1 = \h_2 = \h_3 = (1/2) ( x_1^2 + x_2^2 + x_3^2 + x_4^2 )$,
corresponding to the ``quaternionic oscillator'' (which is an
integrable case \cite{Gdeg,GMspt}). With this, we get
$ p_1 = 4 (c_2 - c_3)$, $p_2 = 4 (c_3 - c_1 )$, $p_3 = 4 (c_1 - c_2 )$;
$q_1 = q_2 = q_3 = 0$.

It is maybe worth pointing out also what happens when only one of the
Hamiltonians, say $\h_1$, is nonzero; this corresponds to a
standard Hamiltonian flow. Setting $\h_1 = H$, $\h_2 = \h_3 = 0$
in the general formulas above, we get
\begin{eqnarray*}
p_1 &=& 0 \ , \ \ p_2 \ = \ c_3 \ \triangle H \ , \ \
p_3 \ = \ - \, c_2 \ \triangle H \ ; \\
q_1 &=& - \ c_2 \( \pa_1^2 H - \pa_2^2 H +\pa_3^2 H -\pa_4^2 H \) \ + \ 2 \, c_3 \( \pa_1 \pa_2 H +\pa_3 \pa_4 H \)  \ , \\
q_2 &=&  - \ c_3 \(\pa_1^2 H - \pa_2^2 H - \pa_3^2 H +\pa_4^2 H \) \ - \ 2 \, c_2 \(\pa_1 \pa_2 H \pa_3 \pa_4 H \) \ , \\
q_3 &=& 2 \, c_2 \( \pa_1 \pa_4 H + \pa_2 \pa_3  H \) \ + \ 2 \,
c_3 \, \( \pa_1 \pa_3 H - \pa_2 \pa_4 H \) \ .
\end{eqnarray*} This shows that even a simple
Hamiltonian flow is in general {\it not} hyperkahler; the
special choice $H = H(x_1^2 + x_2^2 + x_3^2 + x_4^2)$ will of
course produce $q_\alpha = 0$ and hence gives an hyperkahler
flow.


\bigskip\noindent {\bf Acknowledgements.} MAR was
supported by the Spanish Ministry of Science and Innovation under
project FIS2011-22566. GG is supported by the Italian MIUR-PRIN
program under project 2010-JJ4KPA. This article was started and
developed in the course of visits by GG in Universidad Complutense
and by MAR in Universit\`a di Milano; we thank these Institutions
for their support.


\section*{Appendix A. The operator $\P_k (\om )$.}

Let $\om$ be a non-degenerate two-form on the $4n$-dimensional
orientable manifold $M$; and let $\Om$ be the volume form on $M$.
We associate to $\om$ its $2n$-th external power, which we denote
by (from now on $m = 2n$) $\De_m (\om) \in \Lambda^{2m} (M)$:
\begin{equation}
 \De_m (\om ) \ = \ \om \w ... \w \om \ ;
 \end{equation}
being a form of maximal rank on $M$, this is necessarily
proportional to the volume form,
\begin{equation} \De_m (\om) \ = \ p_m (\om) \, \Om \ . \end{equation}
Obviously, the scalar function $p_m (\om) : M \to \R$ is
homogeneous of degree $m$, i.e. $p_m (k \om) = k^{m} p_m (\om )$.
Thus it suffices to study $\De_m (\om )$ on the unit sphere $\Q
\in {\bf Q}$.

We also notice that $\De_m (\om )$ is defined point-wise on $M$;
as discussed in Section \ref{sec:stand_sub}, we can always
transform any hypersymplectic structure to a standard one at any
given point: it is enough to consider $p_m (\om)$ for a standard
quaternionic symplectic structure, i.e. a block reducible one,
spanned by the $\{ \om_\a \}$, or the $\{ \^\om_\a \}$, on each
fundamental block.

We can write $\om$ in coordinates as
\begin{equation}
 \om \ =  \ K_{ij} (x) \ \d x^i \w \d x^j
 \end{equation}
(we will just write $K$ for $K(x)$ in the following); the matrix
$K$ is antisymmetric and of maximal rank. We can then write the
$m$-th external power of $\om$ as
\begin{equation}
 \De_m (\om ) \ = \ (1/m!) \ \epsilon_{i_1 j_1 ... i_m j_m} \
 K_{i_1 j_1} \, ... \, K_{i_m j_m} \ \Om \ := \ p_m (\om ) \, \Om .
 \end{equation}
We will focus on the scalar function $p_m(\om)$, and look at it in
terms of a function defined on the (antisymmetric) matrices $K$
corresponding to $\om$,
\begin{equation}
 \P_m (K) \ := \ \sum_{i_s,j_s=1}^{4n} \ \epsilon_{i_1 j_1 ... i_m j_m} \
K_{i_1 j_1} \, ... \, K_{i_m j_m}  \ .
\end{equation}
This is the function considered in Sections \ref{sec:bkI} and
\ref{sec:canmaps}. The square of $\P_m (K)$ is given by
\begin{equation}
 [\P_m (K ) ]^2 \ = \ \epsilon_{i_1 j_1 ... i_m j_m} \
\epsilon_{a_1 b_1 ... a_m b_m} \ K_{i_1 j_1} \, ... \, K_{i_m j_m}  \
K_{a_1 b_1} \, ... \, K_{a_m b_m} \ .
\end{equation}

We can rewrite $\epsilon_{i_1 j_1 ... i_m j_m} = (-1)^{m/2}
\epsilon_{i_1 ... i_m j_1 ... j_m}$, and the like for
$\epsilon_{a_1 b_1 ... a_m b_m}$:
\begin{equation}
 [\P_m (K ) ]^2 \ = \ \epsilon_{i_1  ... i_m j_1 ... j_m} \
\epsilon_{a_1 ... a_m b_1 ... b_m} \ K_{i_1 j_1} \, ... \, K_{i_m j_m}  \
K_{a_1 b_1} \, ... \, K_{a_m b_m} \ .
\end{equation}
Note that each of the $\epsilon$ symbols depends on $2m$ indices;
hence all the $4n$ coordinates must appear in it. We can then
always operate a permutation in one of them, say the first one, so
that the coordinate indices appear in consecutive order; this will
give a $\pm 1$ sign for the permutation. If we operate the same
permutation also on the indices of the second $\epsilon$ tensor
(thus getting an equal sign which in any case cancels the one
obtained from the previous permutation) we are reduced to an
expression of the type
\begin{equation}
 [\P_m (K ) ]^2 \ = \ \epsilon_{c_1 ...  c_{2m}} \ K_{1 c_1} \, ... \,
 K_{2m, c_{2m}} \ .
 \end{equation}
This is immediately recognized as the determinant of $K$. We have
thus shown that
\begin{equation}
 [\P_m (K)]^2 \ = \ \Det (K) \ ; \ \ \ \P_m (K) \ = \ \pm \, \sqrt{\Det (K) } \ .
 \end{equation}

It follows at once from this that for the product of two matrices we have
\begin{equation}
 \P_m (A B) \ = \ [\pm \, \sqrt{ \Det (A)}]
\ [ \pm \, \sqrt{\Det (B)}] \ = \ \pm \, \sqrt{ \Det (A B)} \ .
\end{equation}
When we consider $\^K = A^T K A$ we thus have
\begin{equation}
 \P_m (A^T K A) \ = \ \P_m (K) \ \Det (A) \ .
 \end{equation}

Similar considerations, up to combinatorial factors, also hold for
$\De_k (\om)$ with $k < m$, and for projections of these to
$2k$-dimensional submanifolds; in particular, to the invariant
four-dimensional subspaces $U_a$.

\section*{Appendix B. Alternative proof of Theorem \ref{sec:hhcan}.1.}

The key step to our proof of Theorem \ref{sec:hhcan}.1 was to show
that $\Theta_{\a \b \ga} = 0$. The vanishing of the $\Theta_{\a \b
\ga}$ depends of course not only on the symmetry of Christoffel
symbols but also on the combinatorial properties of the $K_\a$ and
$Y_\a$. In this Appendix we discuss briefly how these lead to the
vanishing of the $\Theta_{\a \b \ga}$.

%
Let us look separately at the two kinds of terms in
\eqref{eq:thm}. As for those of the form $K^\a_{\ell m}
(Y_\ga)^q_{\ p} (A_i)^p_{\ j} (\d x^i \w \d x^j \w \d x^\ell \w \d
x^m)$, it follows immediately from the symmetry of $(A_i)^p_{\
j}$, and the antisymmetry of $\d x^i \w \d x^j$, that under the
exchange of $i$ and $j$ these change sign, and hence their sum
vanishes.

The other type of terms, i.e. those of the type $K^\a_{\ell m}
(A_i)^q_{\ p}  (Y_\ga)^p_{\ j} (\d x^i \w \d x^j \w \d x^\ell \w
\d x^m)$, require a slightly more careful discussion. Only two
pair of indices $(\ell,m)$ produce nonzero results for the
corresponding element $K^\a_{\ell m}$.

Once we have fixed $\a$ and $\ga$, e.g. $\a = 1$ and $\ga = 3$,
for each element $K^1_{\ell m}$ only elements $(q,j)$ of $Y_3$
with $j$ different from both $\ell$ and $m$ will contribute to
\eqref{eq:thm}. E.g., consider (for $\a=1$, $\ga=3$) the choice
$\ell = 1$, $m = 2$; now only the elements $(Y_3)^1_{\ 3}$ and
$(Y_3)^2_{\ 4}$ satisfy the requirement $(Y_3)^q_{\ j} \not= 0$
for $j\not= 1,2$. Thus, when we remember than now it should also
be $i \not= \ell,m,j$, the only terms actually contributing to
products of the form $K_\a A_i Y_\ga$ will be \beq \label{eq:B1}
K^1_{12} [(A_3)^q_{\ 2} (Y_3)^2_{\ 4} - (A_4)^q_{\ 1} (Y_3)^1_{\
3} ] \ \Om_{(1)} \ . \eeq Exchanging the indices $\ell$ and $m$
will give just the same result. On the other hand, also terms with
$\ell = 3$ and $m=4$ will give a nonzero $K^1_{\ell m}$;
proceeding as above, this will give terms of the type \beq
\label{eq:B2} K^1_{34} [(A_1)^q_{\ 4} (Y_3)^4_{\ 2} - (A_2)^q_{\
3} (Y_3)^3_{\ 1} ] \ \Om_{(1)} \ . \eeq Here again exchanging
$\ell$ and $m$ will give the same result.

If now we sum \eqref{eq:B1} and \eqref{eq:B2}, use $Y_\a = K_\a$
at the reference point, and collect terms using $(A_i)^j_{\ k} =
\Ga^j_{ik} = \Ga^j_{ki}$, we obtain \beq \label{eq:B3} \Ga^q_{14}
\, \( K^1_{34} K^3_{42} - K^1_{12} K^3_{13} \) \ + \ \Ga^q_{23} \,
\( K^1_{12} K^3_{24} - K^1_{34} K^3_{31} \) \ \Om_{(1)} \ . \eeq
Now we observe that, as the $\om_\a$ have the same orientation,
necessarily $K^1_{12}/K^1_{34} = K^3_{24}/K^3_{31}$, hence the
form \eqref{eq:B3} vanishes. The same discussion can be repeated
for other choices of $\a$ and $\ga$, and for negative orientation.


\begin{thebibliography}{49}

\bibitem{AlM} D. V. Alekseevsky and S. Marchiafava,
``Quaternionic structures on a manifold and subordinated
structures," Ann. Mat. Pura Appl. {\bf 171},  205--273 (1996).

\bibitem{Arn} V. I. Arnold, {\it Mathematical Methods
of Classical Mechanics} (Springer, Berlin, 1989).

\bibitem{At1} M. F. Atiyah, {\it Geometry of Yang-Mills fields
(lezioni fermiane)} (SNS, Pisa, 1979).

\bibitem{At2} M. F. Atiyah, ``Hyper-Kahler manifolds'' ({\it Complex geometry and analysis,} Lecture
Notes in Mathematics vol. 1422), ed. V. Villani (Springer, Berlin, 1990).

\bibitem{AtH} M. F. Atiyah and N. J. Hitchin, {\it The
geometry and dynamics of magnetic monopoles.} (Princeton University
Press, Princeton, 1988).

\bibitem{AtMS} M. F. Atiyah, N. S. Manton and  B. J. Schroers,
``Geometric models of matter," Proc. R. Soc. A {\bf 468}, 125--1279, (2012).

\bibitem{Be55} M. Berger,
``Sur les groupes d'holonomie homog\`enes de vari\'et\'es \`a connexion affine et des vari\'et\'es riemanniennes," Bull. Soc. Math. France {\bf 83}, 279--330, (1955).

\bibitem{Cal} E. Calabi, ``Isometric families of
Kahler structures,'' ({\it The Chern symposium 1979}) ed. W. Y.
Hsiang et al. (Springer, New York, 1980)

\bibitem{Cal2} E. Calabi, ``M\'etriques kahleriennes et fibr\'es holomorphes,''
Ann. Sci. E.N.S. {\bf 12}, 269--294, (1979).

\bibitem{Che} S. A. Cherkis, ``Moduli spaces of instantons on
the Taub-NUT space,'' {\it Comm. Math. Phys.} {\bf 290} 719--736, (2009).

\bibitem{ChH} S. A. Cherkis and N. J. Hitchin,  ``Gravitational
Instantons of Type $D_k$,''  Comm. Math. Phys. {\bf 260}, 299--317, (2005).

\bibitem{ChK} S. A. Cherkis and A. Kapustin, ``Hyper-Kahler metrics
from periodic monopoles,'' {\it Phys. Rev. D} {\bf 65}, 084015, (2002).

\bibitem{CCL} S. S. Chern, W. H. Chen and K. S. Lam,
{\it Lectures on Differential Geometry} (World Scientific, Singapore, 2000)

\bibitem{Dan} A. S. Dancer, ``Nahm's equations and hyperkahler geometry,''
Commun. Math. Phys. {\bf 158}, 545--568, (1993).

\bibitem{Dun} M. Dunajski, {\it Solitons, instantons and
twistors} (Oxford University Press, Oxford, 2010)

\bibitem{DuM} M. Dunajski and L. Mason, ``Hyper-Kahler hierarchies
and their twistor theory,''  Commun. Math. Phys. {\bf 213},
641--672 (2003).

\bibitem{EGH} T. Eguchi, P. B. Gilkey  and A. J. Hanson, ``Gravitation, gauge
theories and differential geometry,'' {\it Phys. Rep.} {\bf 66}, 213--393 (1980).

\bibitem{Gdeg} G. Gaeta, ``Quaternionic integrability,'' J. Nonlin. Math. Phys.
 {\bf 18}, 461-474 (2011).

\bibitem{GM} G. Gaeta and P. Morando, ``Hyper-Hamiltonian dynamics,''
{\it J. Phys. A: Math. Gen.} {\bf 35}, 3925--3943, (2002).

\bibitem{GMspt} G. Gaeta and P. Morando, ``Quaternionic integrable systems,''
({\it Symmetry and Perturbation Theory -- SPT2002}) ed.  S. Abenda, G. Gaeta and S. Walcher (World Scientific, Singapore, 2003).

\bibitem{GM2} G. Gaeta and P. Morando, ``A variational principle
for volume-preserving dynamics,'' J. Nonlin. Math. Phys. {\bf 10}, 539--554 (2003).

\bibitem{GR} G. Gaeta and M. A. Rodr\'{\i}guez, ``On the physical applications of
hyper-Hamiltonian dynamics,'' J. Phys. A: Math. Theor. {\bf 41}, 175203 (2008).

\bibitem{GRtnut} G. Gaeta and M. A. Rodr\'{\i}guez, ``Hyperkahler structure of the
Taub-NUT metric,'' J. Nonlin. Math. Phys. {\bf 19}, 1250014, (2012).

\bibitem{GReuclidean} G. Gaeta G and M. A. Rodr\'{\i}guez, ``Structure preserving transformations in hyperkahler Euclidean spaces,'' forthcoming paper.

\bibitem{Hit} N. J. Hitchin, ``The self-duality equations on a Riemann surface,''
Proc. London Math. Soc. {\bf 55}, 59--126 (1987).

\bibitem{HKLR} N. J. Hitchin, A. Karlhede, U. Lindstrom and M. Rocek,
``Hyperkahler metrics and Supersymmetry,'' Commun. Math.
Phys. {\bf 108}, 535--589 (1987).

\bibitem{Jo00} D. D. Joyce, {\it Compact manifolds with special holonomy}
(Oxford Sci. Pub., 2000)

\bibitem{Kir} A. A. Kirillov, {\it Elements of the Theory of
Representations} (Springer, Berlin, 1976)

\bibitem{Man} N. S. Manton, ``A remark on the scattering of BPS monopoles,''
Phys. Lett. B {\bf 110}, 54--56 (1982).

\bibitem{MT} P. Morando and M. Tarallo ``Quaternionic Hamilton equations''
Mod. Phys. Lett. A {\bf 18}, 1841--1847 (2003).

\bibitem{Nak} M. Nakahara, {\it Geometry, Topology and Physics} (IOP, Bristol, 1990)

\bibitem{NS} Ch. Nash and S. Sen, {\it Topology and Geometry for Physicists}
(Academic Press, 1983, reprinted by Dover, London, 2011)

\bibitem{NN} A. Newlander and L. Nirenberg, ``Complex analytic coordinates in almost complex manifolds,'' Ann. Math. {\bf 65}, 391-404 (1957).

\bibitem{NUT} E. Newman, L. Tamburino and T. Unti,
``Empty-space generalization of the Schwarzschild metric,'' J. Math. Phys. {\bf 4}, 915--924 (1963).

\bibitem{PeP} H. Pedersen and Y. S. Poon, ``Hyper-Kahler metrics and a
generalization of the Bogomolny equations,'' Comm. Math.
Phys. {\bf 117}, 569--580, (1988).

\bibitem{Sa89} S. Salamon, {\it Riemannian geometry and holonomy groups}
(Longman, Harlow UK, 1989)

\bibitem{Taub} A. H. Taub, ``Empty space-times admitting a
three parameter group of motions,''  Ann. Math. {\bf 53},
472--490, (1951).

\end{thebibliography}
\end{document}